\documentclass{article}

\usepackage{PRIMEarxiv}

\usepackage[utf8]{inputenc} 
\usepackage[T1]{fontenc}    
\usepackage[hidelinks]{hyperref}       
\usepackage{url}            
\usepackage{booktabs}       
\usepackage{amsfonts}       
\usepackage{nicefrac}       
\usepackage{microtype}      
\usepackage{lipsum}
\usepackage{fancyhdr}       
\usepackage{graphicx}       
\graphicspath{{media/}}     

\usepackage{multirow}
\usepackage{floatrow}
\newfloatcommand{capbtabbox}{table}[][\FBwidth]
\usepackage{blindtext}
\usepackage{balance}
\usepackage{caption}
\usepackage{adjustbox}

\newcommand{\graphrep}{resource-interaction graph\xspace}

\usepackage[dvipsnames]{xcolor}
\usepackage{pgfplots} 
\pgfplotsset{compat=1.17}

\usepackage{subcaption}
\usepackage{orcidlink}

\usepackage{color,soul} 
\usepackage[toc,page]{appendix} 
\usepackage{floatrow}

\definecolor{myblue}{RGB}{68, 119, 170}
\definecolor{mycyan}{RGB}{102, 204, 238}
\definecolor{mygreen}{RGB}{34, 136, 51}
\definecolor{myyellow}{RGB}{204, 187, 68}
\definecolor{myred}{RGB}{238, 102, 119}
\definecolor{mypurple}{RGB}{170, 51, 119}
\definecolor{mygrey}{RGB}{187, 187, 187}

\usepackage{algpseudocode}
\usepackage{algorithm}

\usepackage{listings}
\renewcommand{\figurename}{Fig.}

\usepackage{authblk}

\title{Resource-Interaction Graph: Efficient Graph Representation for Anomaly Detection
}

\author[1]{James Pope\orcidlink{0000-0003-2656-363X}}
\author[2]{Jinyuan Liang\orcidlink{0000-0002-4435-1393}}
\author[4]{Vijay Kumar\orcidlink{0000-0001-5288-0415}}
\author[1]{Francesco Raimondo\orcidlink{0000-0003-2086-7210}}
\author[1]{Xinyi Sun\orcidlink{0000-0001-6545-3607}}
\author[1]{Ryan~McConville\orcidlink{0000-0002-7708-3110}}
\author[2]{\\Thomas Pasquier\orcidlink{0000-0001-6876-1306}}
\author[1]{Rob Piechocki\orcidlink{0000-0002-4879-1206}}
\author[1]{George Oikonomou\orcidlink{0000-0002-1684-6989}}
\author[3]{Bo Luo\orcidlink{0000-0002-2249-8263}}
\author[3]{Dan Howarth\orcidlink{0000-0002-9018-7352}}
\author[4]{Ioannis~Mavromatis\orcidlink{0000-0002-3309-132X}}
\author[4]{\\Adrian~Sanchez-Mompo\orcidlink{0000-0003-0821-4057}}
\author[4]{Pietro Carnelli\orcidlink{0000-0002-4993-5873}}
\author[5]{Theodoros~Spyridopoulos\orcidlink{0000-0001-7575-9909}}
\author[4]{Aftab Khan\orcidlink{0000-0002-3573-6240}}

\affil[1]{University of Bristol, Bristol, UK}
\affil[2]{University of British Columbia, Vancouver, CA}
\affil[3]{Smartia Ltd, Bristol, UK}
\affil[4]{Toshiba Europe Limited, Bristol Research and Innovation Laboratory, Bristol, UK}
\affil[5]{University of Cardiff, Cardiff, UK}

\begin{document}
\maketitle

\begin{abstract}

Security research has concentrated on converting operating system audit logs into suitable graphs, such as provenance graphs, for analysis. However, provenance graphs can grow very large requiring significant computational resources beyond what is necessary for many security tasks and are not feasible for resource constrained environments, such as edge devices.  To address this problem, we present the \textit{resource-interaction graph} that is built directly from the audit log.  We show that the resource-interaction graph's storage requirements are significantly lower than provenance graphs using an open-source data set with two container escape attacks captured from an edge device.  We use a graph autoencoder and graph clustering technique to evaluate the representation for an anomaly detection task.  Both approaches are unsupervised and are thus suitable for detecting zero-day attacks.  The approaches can achieve f1 scores typically over 80\% and in some cases over 90\% for the selected data set and attacks.

\end{abstract}


\section{Introduction}
\label{sec:introduction}


Given recent advances in graph clustering and graph neural network techniques, recent security research has been on representing audit data, such as system call information, as a graph. Process graphs representing the parent-child process relationships are a compact and straightforward graph representation from audit data. However, this approach contains too little information to be useful for many security related tasks as it omits other resources such as users, sockets, and files. Other approaches~\cite{POIROT2019,HOLMES2019} represent numerous nodes\footnote{Node and vertex are used interchangeably throughout this paper.} and edges in the graph, retaining much of the causal and historical information. However, this also means these graphs become very large for typical audit logs exceeding the computational resources of most devices, particularly resource-constrained edge devices. The problem is converting the raw system call auditing data into a graph suitable for security-related classification while preserving resources.

\begin{figure*}[h]
    \centering
    \includegraphics[width=1.0\textwidth]{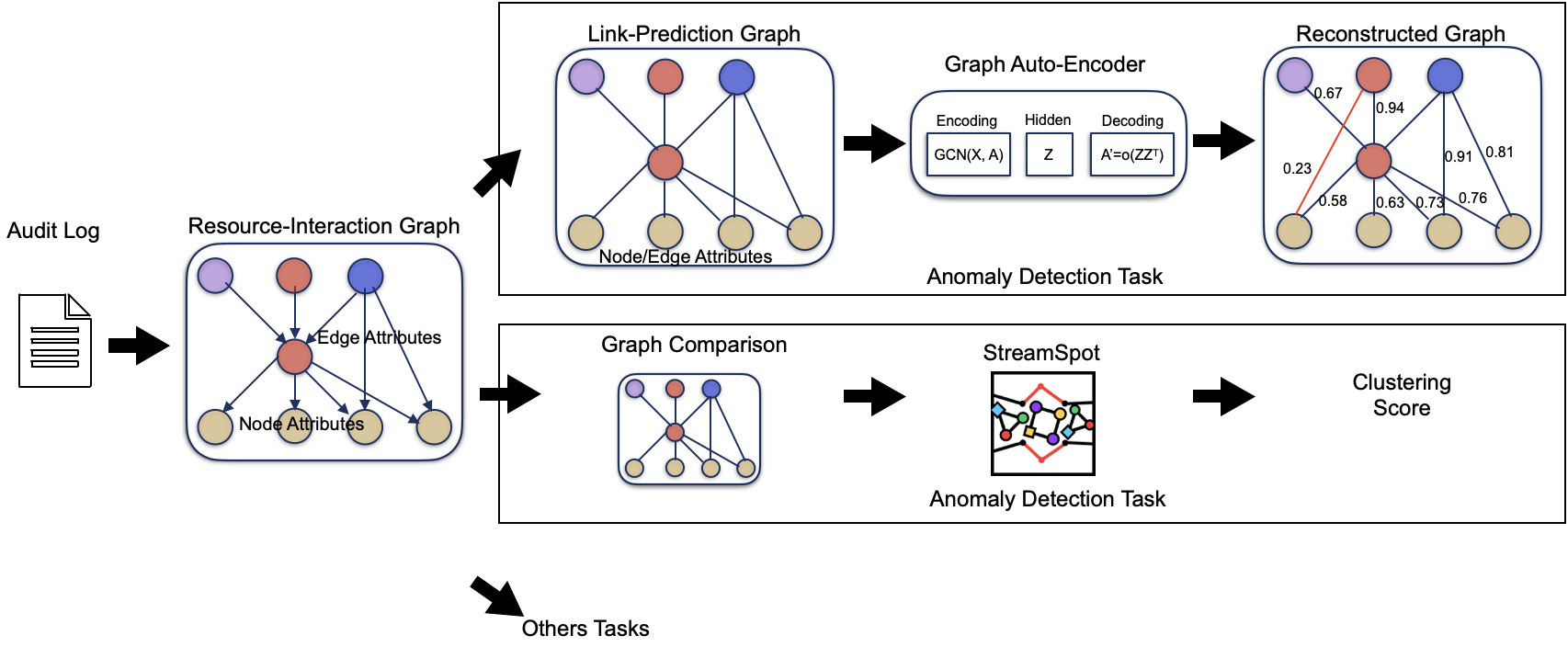}
    \caption{Overview Resource-Interaction Graph and Evaluation}
    \label{fig:overview}
\end{figure*}

In this paper, we present the \textit{resource-interaction graph}, which is a type of spatial-temporal graph. The vertices are computing resources, such as users, processes, and files. Edges capture the interaction between them over time, such as a user creating a process, or a process writing to a file. The vertices represent abstract spatial aspects, while the edges contain time aspects. We select an open source data set collected from a typical edge device to analyse and evaluate the approach.  We show that the resource-interaction graph has 1/3 the number of vertices and 1/10 the number of edges compared to the corresponding provenance graph.  We further show that the number of vertices and the number of edges in the \graphrep tend to grow logarithmically relative to the number of events in the audit log. The amount of temporal information on the edges grows linearly with the number of events, however, this can be mitigated by adjustable time segmentation methods.  

\figurename~\ref{fig:overview} depicts the pipeline that uses the graph representation as input for link prediction and graph clustering anomaly detection. The link prediction approach first converts the \graphrep into a graph where the edges' time attributes are aggregated and combined with node attributes. We then use a graph autoencoder~\cite{VGAE2016} to train on \textit{normal} edges and then predict edges as \{normal, abnormal\}. We also use a recent clustering-based graph anomaly detector~\cite{STREAMSPOT} for evaluation.  Results of both approaches typically achieve over 80\%, confirming that the compact \graphrep representation retains sufficient information for security-related tasks while remaining suitable for resource-constrained devices.  Importantly, both approaches are unsupervised and do not assume any \textit{a priori} threats making them more suitable for detecting zero-day attacks. 


The contribution of the paper is an efficient graph representation of resources and time information for audit logs (e.g. auditd). We empirically show that the graph size tends to be bounded by a logarithmic function for typical logs and a fraction of the equivalent provenance graph.  The analysis source code, along with the data set, is provided as open-source~\cite{CONTAINER_ESCAPE}.

The paper is structured as follows: Section \autoref{sec:dataset} briefly describes the data set, section \autoref{sec:graph} presents the resource-interaction graph, section \autoref{sec:analysis} analyses the graph's growth, section \autoref{sec:eval} details the evaluation and results followed by related (\autoref{sec:related_work}) and future (\autoref{sec:future_work}) work, and section \autoref{sec:conclusion} concludes our work.

\section{Container Escape Dataset}
\label{sec:dataset}

For analysis and evaluation, we use a previous data set~\cite{POPE2021,CONTAINER_ESCAPE} collected from a resource-constrained Linux edge device and similarly configured virtual machine.   The data set contains a denial of service (\textit{DoS}) and privilege escalation attack (\textit{Privesc}) related to container escape.  There are 64 denial of service attacks and 64 privilege escalation attacks with the edge device and a simulated virtual machine for a total of 256 audit logs.  The edge device used to collect the data was the UMBRELLA node~\cite{UMBRELLA_NODE}, which uses a 32-bit processor.  UMBRELLA \cite{UMBRELLA} is an open, programmable, smart city and industrial internet of things (IoT) testbed.  Throughout this paper we abbreviate the edge device as \textit{UM} and the virtual machine as \textit{VM}.

The denial of service attack mounts a filesystem, external to the container, and writes to a shell script.  The attack then uses the cgroup notify/release mechanism to execute the shell on the host.  The privilege escalation attack uses the container volume to mount a file system over the host file system.  The attack then writes to a permissions file modifying the user's privileges.  Both attacks are launched from within a container.

Each audit log covers 15 minutes where three containers are running, and an attack is launched at some random time. The time and duration of each attack are known and provided with each audit log. The individual events are not labelled but, instead, time intervals are used.  All events that occur during the attack are annotated anomalies, meaning that normal events and those related to the attack are labelled as anomalies. To help mitigate this, we only annotate an edge as an anomaly when the event creates a new edge, which means that edges labelled as \textit{normal} are correct. Edges labelled as \textit{abnormal} may be the result of normal or abnormal events.






\section{Resource-Interaction Graph}
\label{sec:graph}
In this section, we first describe an exemplary audit event. We then discuss how time information has traditionally been aggregated using time intervals as a prelude to how this can also be applied to the time information stored on the \graphrep's edges. We then provide details on how events are converted into a graph, followed by an analysis of the graph size.


\subsection{Audit Event}
\label{sec:audit-event}

This section describes how audit logs are converted into the audit graph. The \textit{auditd} log files consist of events comprised of a set of records which themselves are comprised of key-value pairs. Each event has an \textit{event type}, and most of the events ($\sim$ 97\%) are \textit{syscall} events (we currently ignore the other event types). For instance, \autoref{lstevent} provides a sample log entry from a privilege escalation attack. Next, we describe a representation of the events using a \textit{non-graph time representation}.

\begin{lstlisting}[label=lstevent,caption=Privilege Escalation Audit Event,frame=tb]
type=SYSCALL msg=audit(1632851805.333:76118):
    syscall=59 ppid=12261 pid=12272 uid=0 comm="escape.sh" exe="/bin/busybox"
type=EXECVE: argc=2 a0="/bin/sh" a1="/escape.sh"
type=CWD: cwd="/privesc"
type=PATH: name="/escape.sh" inode=667188
type=PATH: name="/bin/sh" inode=65711
type=PATH: name="/lib/ld-musl-x86_64.so.1" inode=65873
type=PROCTITLE: proctitle=72756E6300696E6974
\end{lstlisting}

\subsection{Non-graph Time Segmentation Event Representation}
\label{sec:time-segmentation}

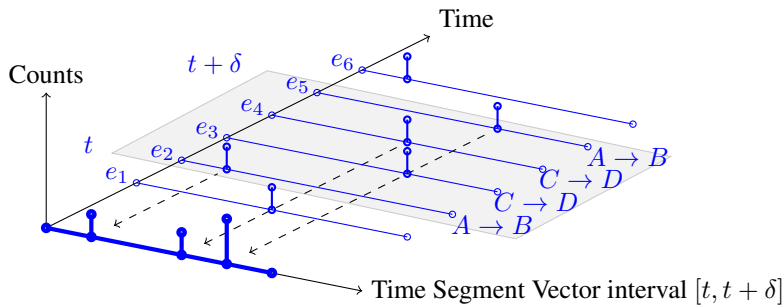
\begin{figure}
    \centering


\begin{tikzpicture}[x={(1cm,0.5cm)},z={(0cm,1cm)},y={(1cm,-0.2cm)}, scale=0.6]

    {
    \usetikzlibrary{shapes.geometric}
    \tikzstyle{every node}=[trapezium, draw,
trapezium right angle=140, trapezium left angle=40]
    \node[minimum height=1.5cm, minimum width=1.0cm, rotate=-12, fill=gray!10, text opacity=0, draw opacity=0.2]
    at (4.50, 3.15, 0) {.....};
    }

    \draw[->] (0,0,0) --++ (8.5,0,0) node[above right] {Time};
    \draw[->] (0,0,0) --++ (0,7,0) node[right] {Time Segment Vector interval $[t,t+\delta]$};
    \draw[->] (0,0,0) --++ (0,0,3) node[above] {Counts};

    
    
    Add dashed line as visual guide to how counts are computed
    \draw [<-, dashed] (0.5,3,0) --++ (4.3,0,0);
    \draw [<-, dashed] (0.5,1,0) --++ (2.3,0,0); 
    \draw [<-, dashed] (0.5,4,0) --++ (5.3,0,0); 
    

    \draw[blue] (3.00,-1.75,0) node [left]{$t$}; 
    \draw[blue] (6.50,-1.95,0) node [left]{$t+\delta$}; 
    
    \draw[blue] plot[mark=o] coordinates {(2,0,0) (2,6,0)};
    \draw[blue] (2+0.25,0.0-0.20,0) node [left]{$e_{1}$}; 
    
    \draw[blue] plot[mark=o] coordinates {(3,0,0) (3,6,0)};
    \draw[blue] (3+0.25,0.0-0.20,0) node [left]{$e_{2}$}; 
    
    \draw[blue] plot[mark=o] coordinates {(4,0,0) (4,6,0)};
    \draw[blue] (4+0.25,0.0-0.20,0) node [left]{$e_{3}$}; 
    
    \draw[blue] plot[mark=o] coordinates {(5,0,0) (5,6,0)};
    \draw[blue] (5+0.25,0.0-0.20,0) node [left]{$e_{4}$}; 
    
    \draw[blue] plot[mark=o] coordinates {(6,0,0) (6,6,0)};
    \draw[blue] (6+0.25,0.0-0.20,0) node [left]{$e_{5}$}; 
    
    \draw[blue] plot[mark=o] coordinates {(7,0,0) (7,6,0)};
    \draw[blue] (7+0.25,0.0-0.20,0) node [left]{$e_{6}$}; 

    
    \draw[blue] (3.25,7.75,0) node [left]{$A \to B$}; 
    \draw[blue] (4.25,7.75,0) node [left]{$C \to D$}; 
    \draw[blue] (5.25,7.75,0) node [left]{$C \to D$}; 
    \draw[blue] (6.25,7.75,0) node [left]{$A \to B$}; 
    
    \draw[blue, thick] plot[mark=o] coordinates {(7,1,0) (7,1,0.5)};
    \draw[blue, thick] plot[mark=o] coordinates {(2,3,0) (2,3,0.5)};
    \draw[blue, thick] plot[mark=o] coordinates {(3,1,0) (3,1,0.5)};
    \draw[blue, thick] plot[mark=o] coordinates {(4,4,0) (4,4,0.5)};
    \draw[blue, thick] plot[mark=o] coordinates {(5,3,0) (5,3,0.5)};
    \draw[blue, thick] plot[mark=o] coordinates {(6,4,0) (6,4,0.5)};

    \draw[blue, ultra thick] plot[mark=o] coordinates {(0,0,0) (0,5,0)};
    \draw[blue, ultra thick] plot[mark=o] coordinates {(0,1,0) (0,1,0.5)};
    \draw[blue, ultra thick] plot[mark=o] coordinates {(0,3,0) (0,3,0.5)};
    \draw[blue, ultra thick] plot[mark=o] coordinates {(0,4,0) (0,4,1.0)};

\end{tikzpicture}
    \caption{Time segmentation representation}%
  \label{fig:audit-time}
\end{figure}

Audit logs can also be described in different representations based on time segmentation~\cite{KUBANOMALY}. The time segmentation approach assumes independence between vectors and is relatively different from other time series approaches (e.g. recurrent neural networks). First, all the events in the audit log are sorted by their timestamp. Then given some time interval $[t,t+\delta]$, where $\delta$ is a hyper-parameter, the system call type is one-hot encoded and summed into a vector that we denote as the \textit{time segment vector}, to represent the time interval. \figurename~\ref{fig:audit-time} depicts how four events are aggregated over the time interval $[t,t+\delta]$. Events $e_2$ and $e_5$ involve resources $A$ and $B$ while events $e_3$ and $e_4$ involve resources $C$ and $D$. The resulting time segment vectors can be used to train an auto-encoder to recognise \textit{normal} from \textit{abnormal} vectors. It is important to mention that this mixing of events loses the resource interaction information. The length of a time segment vector is determined by the maximum system call type, which differs depending on the machine, as Table \ref{tab:syscall-types} shows, typically no more than 400.

\begin{table}
    \centering
    \begin{tabular}{l|c}
    \hline
    System Call Name  &  System Call Index  \\
    \hline
    \_\_NR\_restart\_syscall &  0 \\
    \_\_NR\_exit &  1   \\
    \_\_NR\_fork &  2   \\
    \_\_NR\_read &  3 \\
    \_\_NR\_write &  4 \\
    \_\_NR\_open & 5 \\
    \_\_NR\_execve &  11  \\
    \_\_NR\_chmod & 15 \\
    ... & ... \\
    \_\_NR\_rseq & 398 \\
    \hline
    \end{tabular}
    \caption{System call types\\(from unistd-common.h)}%
    \label{tab:syscall-types}
\end{table}

The anomalous events can fall across the boundary of a time segment and may be harder to detect. The time intervals can be arranged to overlap with each other to somewhat address this issue.  Even with this lossy representation, time segmentation approaches can be effective at detecting anomalies. A significant advantage of this approach is a compact representation. When an anomaly is detected, it also has the benefit of capturing roughly when the anomaly occurred. However, it lacks any resource information about the anomaly. A serious disadvantage of this approach is that it is difficult to explain what resources were involved to a security analyst. We use a similar approach for aggregating the events over time, however, this is done per edge in the graph instead of the entire set of audit events.




\subsection{Edge Attribute Event Representation}
\label{sec:edge-attributes}

\begin{figure}[t]
    \centering
    \includegraphics[width=0.5\textwidth]{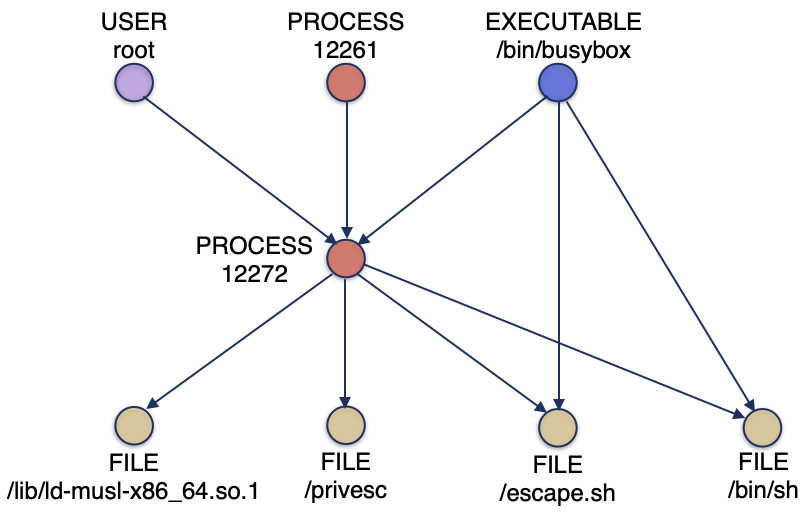}
    \caption{Example Event to Graph}%
    \label{fig:privesc-event}
\end{figure}

\figurename~\ref{fig:privesc-event} presents a sample process tree graph converted from the audit event presented in Listing \ref{lstevent}. The graph represents an instantaneous event and does not include a time component. For subsequent events, existing edges are updated by appending to the edge's attributes with the (timestamp, system call type) tuple, similar to time segmentation. Subsequent interaction between the nodes appends another entry to the edge attributes. \figurename~\ref{fig:edge-time} depicts how the events in \figurename~\ref{fig:audit-time} would be represented in the edges of the resource-interaction graph.

\begin{table}
    \centering
    \begin{tabular}{l|c}
    \hline
    Node Type  &  Audit Key(s)  \\
    \hline
    Process &  pid, ppid \\
    Executable &  exe   \\
    User &  uid     \\
    File &  cwd, path, filename     \\
    Socket &  address     \\
    \hline
    \end{tabular}
    \caption{Node Types}%
  \label{tab:node-types}
\end{table}

\begin{figure}
  \begin{subfigure}[t]{.48\textwidth}
    \centering


\begin{tikzpicture}[x={(1cm,0.5cm)},z={(0cm,1cm)},y={(1cm,-0.2cm)}, scale=0.40]

    \draw[->] (0,0,0) --++ (8.5,0,0) node[above right] {Time};
    \draw[->] (0,0,0) --++ (0,7,0) node[right] {Edge Vector};
    \draw[->] (0,0,0) --++ (0,0,3) node[above] {Counts};

    
    
    \draw [<-, dashed] (0.5,1,0) --++ (2.3,0,0); 
    \draw [<-, dashed] (0.5,4,0) --++ (5.3,0,0); 
    
    \draw[blue] (3.15,7.75,0) node [left]{$A \to B$}; 
    \draw[blue] (6.15,7.75,0) node [left]{$A \to B$}; 
    
    
    
    
    \draw[blue] plot[mark=o] coordinates {(3,0,0) (3,6,0)};
    \draw[blue] (3+0.25,0.0-0.20,0) node [left]{$e_{2}$}; 
    
    
    
    \draw[blue] plot[mark=o] coordinates {(6,0,0) (6,6,0)};
    \draw[blue] (6+0.25,0.0-0.20,0) node [left]{$e_{5}$}; 
    

    \draw[blue, thick] plot[mark=o] coordinates {(3,1,0) (3,1,0.5)};
    \draw[blue, thick] plot[mark=o] coordinates {(6,4,0) (6,4,0.5)};

    \draw[blue, ultra thick] plot[mark=o] coordinates {(0,0,0) (0,5,0)};
    \draw[blue, ultra thick] plot[mark=o] coordinates {(0,1,0) (0,1,0.5)};
    \draw[blue, ultra thick] plot[mark=o] coordinates {(0,4,0) (0,4,0.5)};

\end{tikzpicture}
    \caption{Edge A to B}
    \label{fig:edge-time-left}
  \end{subfigure}
  \hfill
  \begin{subfigure}[t]{.48\textwidth}
    \centering


\begin{tikzpicture}[x={(1cm,0.5cm)},z={(0cm,1cm)},y={(1cm,-0.2cm)}, scale=0.40]

    \draw[->] (0,0,0) --++ (8.5,0,0) node[above right] {Time};
    \draw[->] (0,0,0) --++ (0,7,0) node[right] {Edge Vector};
    \draw[->] (0,0,0) --++ (0,0,3) node[above] {Counts};

    
    
    \draw [<-, dashed] (0.5,3,0) --++ (4.3,0,0);
    \draw [<-, dashed] (0.5,4,0) --++ (3.8,0,0); 
    
    \draw[blue] (4.0,8.5,0) node [left]{$C \to D$}; 
    \draw[blue] (5.3,8.5,0) node [left]{$C \to D$}; 
    
    
    
    
    
    \draw[blue] plot[mark=o] coordinates {(4,0,0) (4,6,0)};
    \draw[blue] (4+0.25,0.0-0.20,0) node [left]{$e_{3}$}; 
    
    \draw[blue] plot[mark=o] coordinates {(5,0,0) (5,6,0)};
    \draw[blue] (5+0.25,0.0-0.20,0) node [left]{$e_{4}$}; 
    
    

    \draw[blue, thick] plot[mark=o] coordinates {(4,4,0) (4,4,0.5)};
    \draw[blue, thick] plot[mark=o] coordinates {(5,3,0) (5,3,0.5)};

    \draw[blue, ultra thick] plot[mark=o] coordinates {(0,0,0) (0,5,0)};
    \draw[blue, ultra thick] plot[mark=o] coordinates {(0,3,0) (0,3,0.5)};
    \draw[blue, ultra thick] plot[mark=o] coordinates {(0,4,0) (0,4,0.5)};

\end{tikzpicture}
    \caption{Edge C to D}
    \label{fig:edge-time-right}
  \end{subfigure}

\caption{Edge time representation }
\label{fig:edge-time}
\end{figure}

We make several observations referencing non-graph time segmentation. The resource-interaction graph can be viewed as first partitioning all the events into their respective edges and then representing a time segment vector. We could introduce multiple time intervals for the edge events resulting in multiple time segment vectors per edge. Alternatively, we could consider an infinite time interval and keep the counts on a one-time segment vector for each edge, avoiding the cost of storing the timestamp and system call type tuple.

\subsection{Audit Event to Graph Conversion}

Algorithm \ref{alg:event2graph} describes in detail how events update a graph.  The inputs are a graph $G$, the event $e$, and a boolean value $build\_tree$.  The result is either a graph that includes the process tree or a graph with pseudo process nodes, denoted respectively as the \textit{process tree} graph and \textit{pseudo process} graph.  In both cases, the resulting graph is directed with no cycles.

\begin{algorithm}
\caption{Audit Event to Graph Conversion}
\label{alg:event2graph}
\begin{algorithmic}[1]
\State\textbf{Input: Graph \textit{G}, Event \textit{e}, boolean \textit{build\_tree} }
\State \# The exe can coincide with a path so add prefix to make unique
\State \# This also helps avoid cycles
\State $e.exe \gets 'executable:' + e.exe$

\If{$not$ $build\_tree$}
    \State \#Concatenate uid and exe to make pid
    \State $e.pid \gets e.uid + e.exe$.
    \label{line:pseudo}
\EndIf

\State $G.addUpdateNode(e.pid, node\_type='PROCESS')$
\State $G.addUpdateNode(e.exe, node\_type='EXECUTABLE')$
\State $G.addUpdateNode(e.uid, node\_type='USER')$
\State $G.addUpdateEdge(e.uid, e.pid, edge\_type=e.syscall, e.timestamp)$

\If{$build\_tree$}
\label{line:proc_tree}
    \State \#Add edge from parent to child process
    \State $G.addUpdateNode(e.ppid, node\_type='PROCESS')$
    \State $G.addUpdateEdge(e.ppid, e.pid, edge\_type=e.syscall, e.timestamp)$ 
\EndIf


\State \# Handle socket records
\If{$e.hasSockaddr()$}
    \State \# Address is formatted based on family
    \State $G.addUpdateNode(e.address, node\_type='SOCKET')$
    \State $G.addUpdateEdge(e.pid, e.address, edge\_type=e.syscall, e.timestamp)$ 
\EndIf

\State \# Handle path records (may have more than one)
\For{$path$ in $e.paths()$}
    \If{$len(path.name) > 0 $}
        \State $G.addUpdateNode(path.name, node\_type='FILE')$
        \State $G.addUpdateEdge(e.pid, path.name, edge\_type=e.syscall, e.timestamp)$ 
    \EndIf
\EndFor

\State \# Handle execve records
\If{$e.hasExecve()$}
    \For{$arg$ in $e.arguments()$}
        \If{$arg.startswith('/')$}
            \State $G.addUpdateNode(arg, node\_type='FILE')$
            \State $G.addUpdateEdge(e.pid, arg, edge\_type=e.syscall, e.timestamp)$ 
        \EndIf
    \EndFor
\EndIf

\end{algorithmic}

\end{algorithm}

\begin{figure}[t]
  \begin{subfigure}[t]{.47\textwidth}
    \centering
    \includegraphics[width=\linewidth]{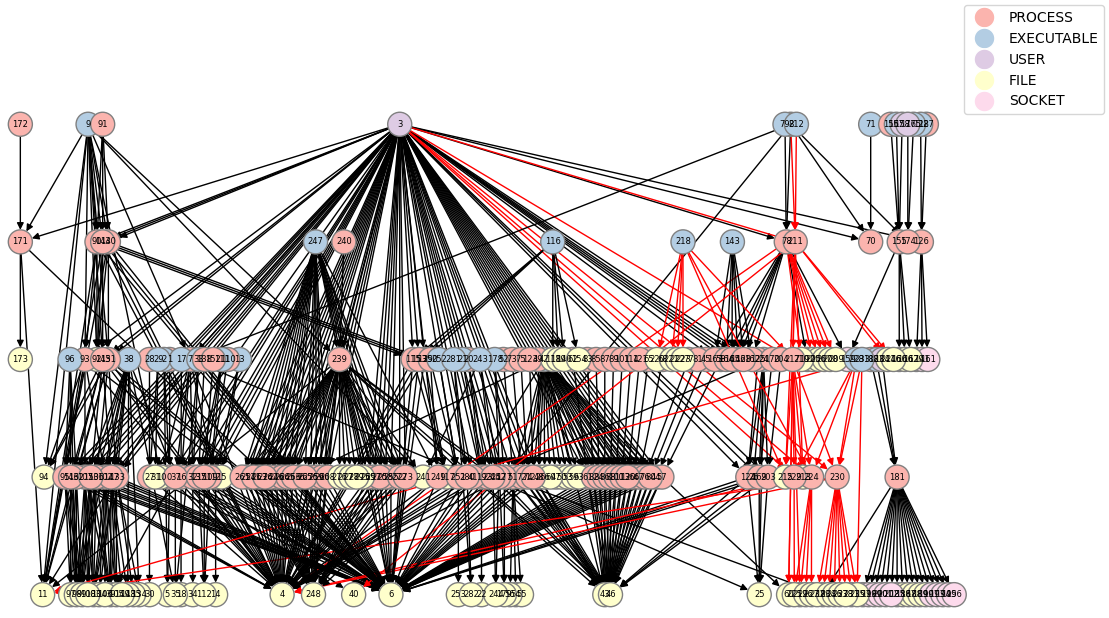}
    \caption{Virtual Machine, Process Tree}
    \label{fig:graph-tree-privesc-vm}
  \end{subfigure}
  \hfill
  \begin{subfigure}[t]{.47\textwidth}
    \centering
    \includegraphics[width=\linewidth]{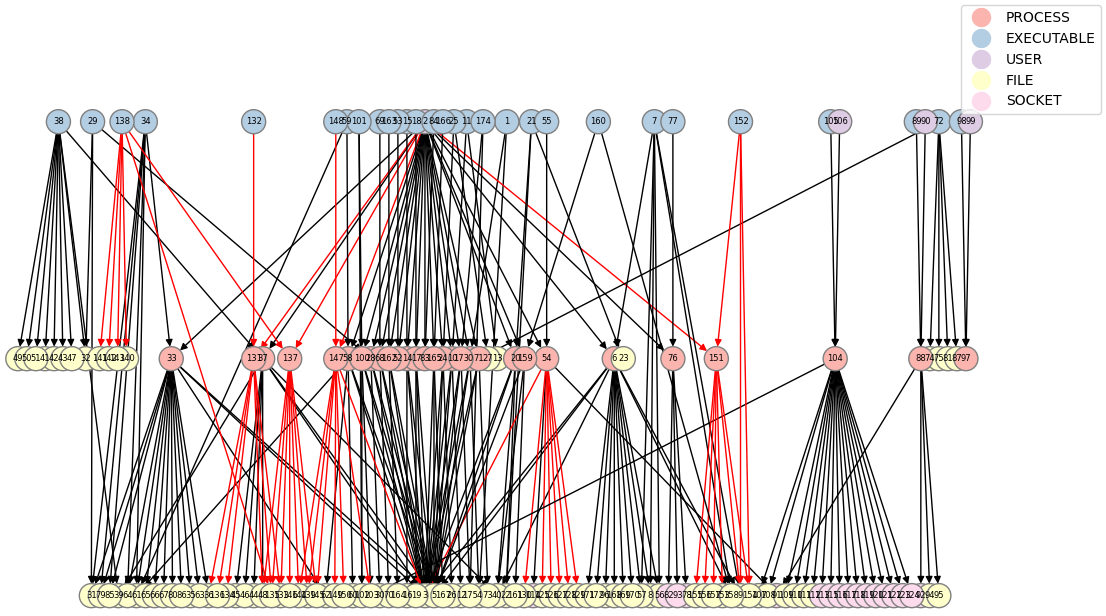}
    \caption{Virtual Machine, Pseudo Process}
    \label{fig:graph-pseudo-privesc-vm}
  \end{subfigure}
  \hfill
  \begin{subfigure}[t]{.47\textwidth}
    \centering
    \includegraphics[width=\linewidth]{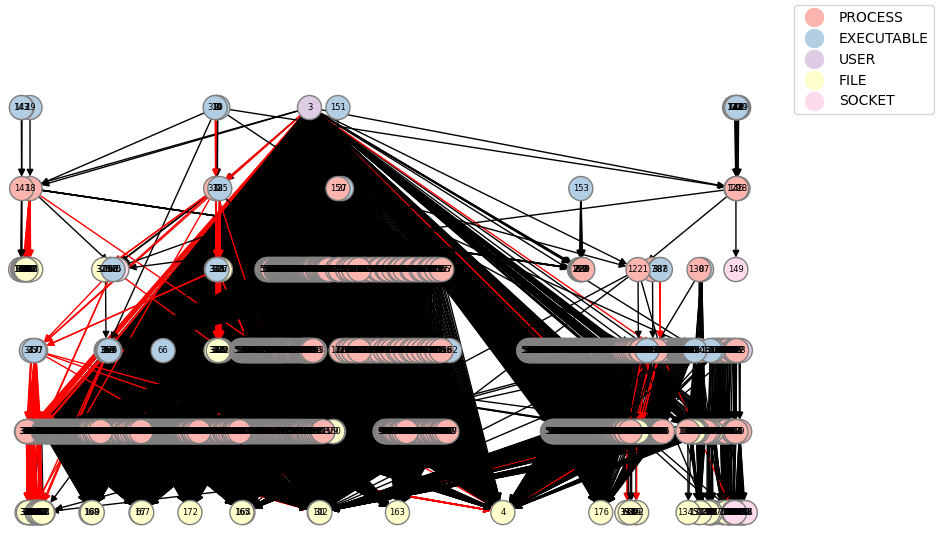}
    \caption{Edge Device, Process Tree}
    \label{fig:graph-tree-privesc-um}
  \end{subfigure}
  \hfill
  \begin{subfigure}[t]{.47\textwidth}
    \centering
    \includegraphics[width=\linewidth]{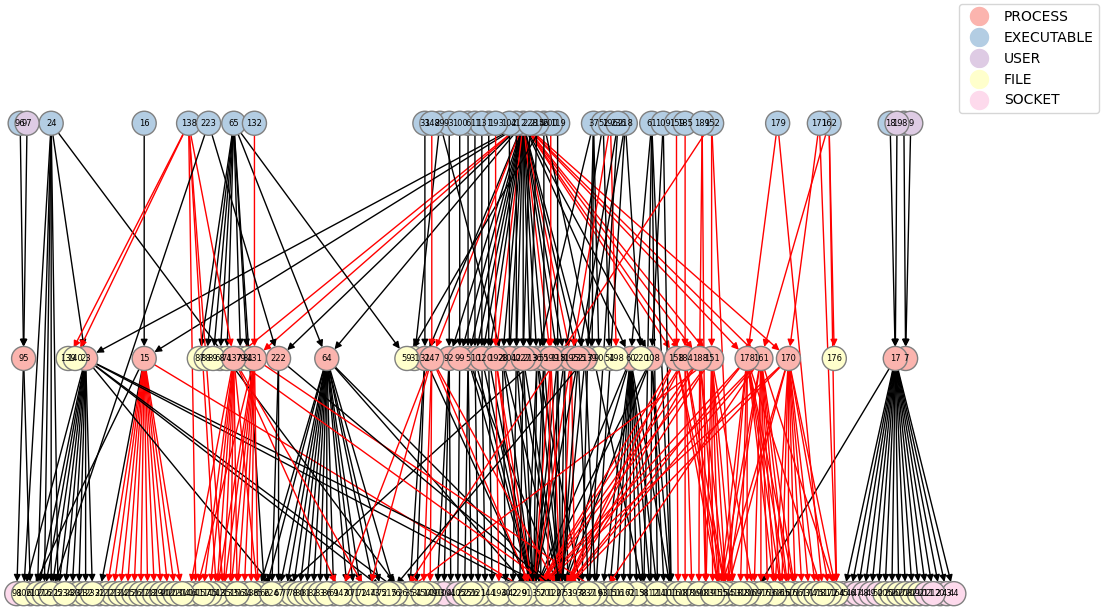}
    \caption{Edge Device, Pseudo Process}
    \label{fig:graph-pseudo-privesc-um}
  \end{subfigure}

\caption{Example Graphs}
\label{fig:graph}
\end{figure}

The graph function \textit{G.addNode(identifier, node\_type)} will add, if the node does not already exist, the node to the graph with the specified $node\_type$, listed in Table \ref{tab:node-types}.  The graph function \textit{G.addUpdateEdge(from, to, edge\_type, timestamp)} will update, or add, the edge between the $from$ and $to$ identifiers with the specified $timestamp$ and $syscall$, listed in Table \ref{tab:syscall-types}.  The event $e$ has all the key/value pairs, and we assume these can be retrieved with object-attribute notation (e.g. $e.timestamp$). On Line \ref{line:proc_tree}, when $build\_tree$ is $true$, the process tree, along with the other resources, is converted into the graph. When $build\_tree$ is $false$ we collapse the process tree into one $pseudo$ process where all process identifiers between a user identifier and executable are grouped into one process node. This occurs on Line \ref{line:pseudo}. This effectively restricts the length of any path to just two as shown in Figures \ref{fig:graph-pseudo-privesc-vm} and \ref{fig:graph-pseudo-privesc-um}.  This furthermore eliminates all the intermediate process nodes and associated edges, greatly reducing the total number of nodes and edges in the graph, as can be seen in \figurename~\ref{fig:graph}. To be clear, the pseudo-process conversion only deals with one event at a time requiring minimal space (i.e. we are not explicitly creating the process tree and then collapsing).

For subsequent events involving the same set of nodes, the edge attributes are appended with the new system call information. If a node already exists, no update occurs. Thus nodes are only created the first time the resource is used in an event. Edges are only created the first time an event results in an interaction between nodes. Finally, edge attributes are updated for each event.


\section{Graph Growth Analysis}
\label{sec:analysis}

In this section we examine the growth of graphs using the number of nodes and edges as an indication of the storage and computational costs. We compare the growth of \graphrep with provenance graphs.  Then we provide empirical evidence that the \graphrep growth is logarithmic.

\subsection{Graph Growth Comparison with Provenance Graphs}
\label{sec:provenance-comparison}

\begin{figure}[t]
    \centering
    \includegraphics[width=\linewidth]{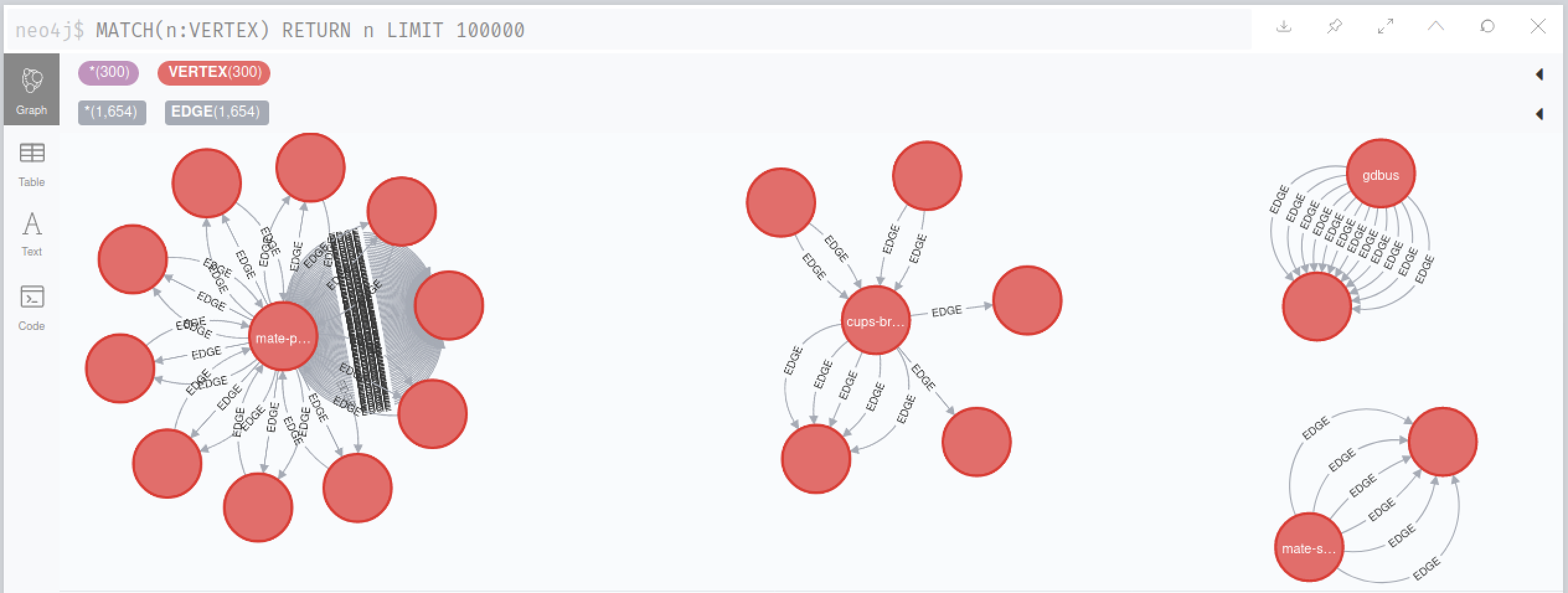}
    \caption{Provenance Graph Example}
    \label{fig:prov_graph}
\end{figure}

We compare the growth of \graphrep  variants with provenance graphs.  The provenance graphs are built using the SPADE tool \cite{gehani2012spade} based on the Open Provenance Model specification \cite{TRACE2021,OPM2011}.  Unfortunately the SPADE tool only supports auditd logs from 64-bit machines (the mapping of syscall number to a name is based on the 64-bit header file).  Therefore, for this section only, the results are based on a 64-bit virtual machine where we repeated the denial of service attack described in the container escape dataset \cite{CONTAINER_ESCAPE}.  \figurename~\ref{fig:prov_graph} shows a subset of the provenance graph.  The provenance graph is directed and allows multiple edges between vertices.  Similar to the \graphrep, the provenance graph vertices represent resources and the edges represent one of several different relationships between vertices.

\begin{table}
\centering
\caption{Example Provenance and Resource Interaction Graph Comparison}
\label{tab:prov-ri-comparison}
\begin{tabular}[t]{lcc}
\toprule
&\# Vertices & \# Edges\\
\midrule
Provenance Graphs & 535 & 2121 \\
Process Tree (RI) Graph & 215 & 377 \\
Pseudo Process (RI) Graph & 172 & 213 \\
\bottomrule
\end{tabular}
\end{table}%

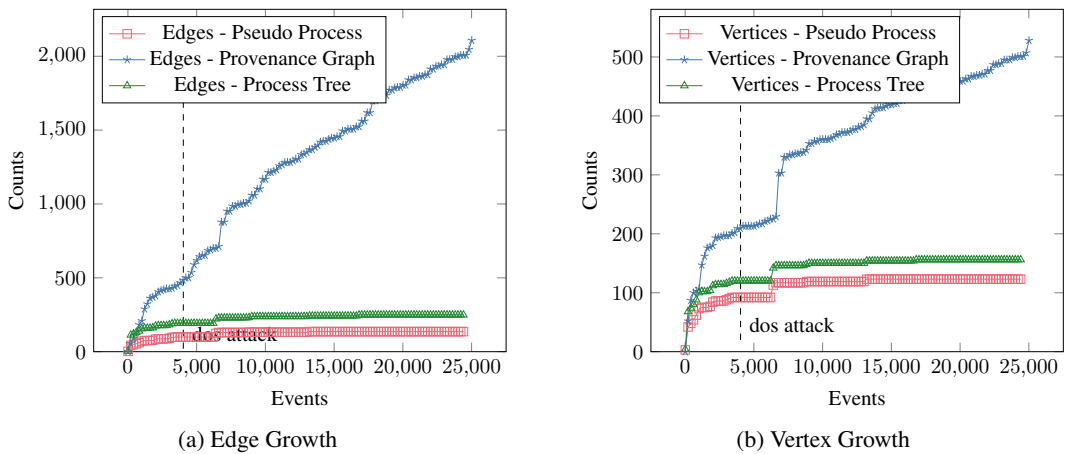
\begin{figure}

  \begin{subfigure}[t]{.45\textwidth}
    \centering
\begin{tikzpicture}[scale=0.80]
\definecolor{RED}{RGB}{238, 102, 119}
\definecolor{BLUE}{RGB}{68, 119, 170}
\definecolor{GREEN}{RGB}{34, 136, 51}
\definecolor{PURPLE}{RGB}{170, 51, 119}
\definecolor{CYAN}{RGB}{102, 204, 238}
\definecolor{YELLOW}{RGB}{204, 187, 68}
\definecolor{GREY}{RGB}{187, 187, 187}
    \begin{axis}[
        xlabel=Events,
        ylabel=Counts,
        ymin=0,
        ymax=2317.7000000000003,
        scaled y ticks = false,
        scaled x ticks = false,
        y tick label style={/pgf/number format/.cd, fixed, fixed zerofill,precision=0},
        x tick label style={/pgf/number format/.cd, fixed, fixed zerofill,precision=0},
        legend style={at={(0.02,0.98)},anchor=north west }]
\addplot[sharp plot,mark=square,RED, error bars/.cd, y dir=both, y explicit] plot coordinates
  {
(10,2)
(210,42)
(410,48)
(610,54)
(810,61)
(1010,71)
(1210,75)
(1410,75)
(1610,75)
(1810,77)
(2010,85)
(2210,87)
(2410,87)
(2610,87)
(2810,87)
(3010,91)
(3210,97)
(3410,99)
(3610,99)
(3810,99)
(4010,99)
(4210,99)
(4410,99)
(4610,99)
(4810,99)
(5010,99)
(5210,99)
(5410,99)
(5610,99)
(5810,99)
(6010,99)
(6210,99)
(6410,124)
(6610,130)
(6810,130)
(7010,130)
(7210,130)
(7410,130)
(7610,130)
(7810,130)
(8010,130)
(8210,130)
(8410,130)
(8610,130)
(8810,132)
(9010,132)
(9210,132)
(9410,132)
(9610,132)
(9810,132)
(10010,132)
(10210,132)
(10410,132)
(10610,132)
(10810,132)
(11010,132)
(11210,132)
(11410,132)
(11610,132)
(11810,132)
(12010,132)
(12210,132)
(12410,132)
(12610,132)
(12810,132)
(13010,132)
(13210,136)
(13410,136)
(13610,136)
(13810,136)
(14010,136)
(14210,136)
(14410,136)
(14610,136)
(14810,136)
(15010,136)
(15210,136)
(15410,136)
(15610,136)
(15810,136)
(16010,136)
(16210,136)
(16410,136)
(16610,136)
(16810,136)
(17010,136)
(17210,136)
(17410,136)
(17610,136)
(17810,136)
(18010,136)
(18210,136)
(18410,136)
(18610,136)
(18810,136)
(19010,136)
(19210,136)
(19410,136)
(19610,136)
(19810,136)
(20010,136)
(20210,136)
(20410,136)
(20610,136)
(20810,136)
(21010,136)
(21210,136)
(21410,136)
(21610,136)
(21810,136)
(22010,136)
(22210,136)
(22410,136)
(22610,136)
(22810,136)
(23010,136)
(23210,136)
(23410,136)
(23610,136)
(23810,136)
(24010,136)
(24210,136)
(24410,136)
  };
\addlegendentry{Edges - Pseudo Process}
\addplot[sharp plot,mark=star,BLUE, error bars/.cd, y dir=both, y explicit] plot coordinates
  {
(10,0)
(210,55)
(410,90)
(610,131)
(810,182)
(1010,209)
(1210,289)
(1410,323)
(1610,362)
(1810,370)
(2010,378)
(2210,405)
(2410,415)
(2610,420)
(2810,426)
(3010,426)
(3210,430)
(3410,442)
(3610,449)
(3810,467)
(4010,484)
(4210,500)
(4410,500)
(4610,536)
(4810,589)
(5010,617)
(5210,644)
(5410,652)
(5610,652)
(5810,684)
(6010,689)
(6210,700)
(6410,700)
(6610,712)
(6810,877)
(7010,877)
(7210,951)
(7410,951)
(7610,985)
(7810,985)
(8010,997)
(8210,997)
(8410,1005)
(8610,1005)
(8810,1021)
(9010,1059)
(9210,1059)
(9410,1102)
(9610,1102)
(9810,1168)
(10010,1168)
(10210,1213)
(10410,1213)
(10610,1223)
(10810,1231)
(11010,1258)
(11210,1266)
(11410,1280)
(11610,1282)
(11810,1282)
(12010,1292)
(12210,1304)
(12410,1304)
(12610,1334)
(12810,1339)
(13010,1349)
(13210,1368)
(13410,1368)
(13610,1384)
(13810,1396)
(14010,1419)
(14210,1425)
(14410,1425)
(14610,1438)
(14810,1446)
(15010,1446)
(15210,1456)
(15410,1456)
(15610,1494)
(15810,1494)
(16010,1507)
(16210,1507)
(16410,1507)
(16610,1523)
(16810,1523)
(17010,1561)
(17210,1561)
(17410,1618)
(17610,1618)
(17810,1698)
(18010,1698)
(18210,1720)
(18410,1720)
(18610,1735)
(18810,1735)
(19010,1771)
(19210,1771)
(19410,1786)
(19610,1790)
(19810,1790)
(20010,1805)
(20210,1805)
(20410,1840)
(20610,1840)
(20810,1855)
(21010,1855)
(21210,1865)
(21410,1865)
(21610,1875)
(21810,1875)
(22010,1913)
(22210,1913)
(22410,1933)
(22610,1933)
(22810,1943)
(23010,1943)
(23210,1977)
(23410,1979)
(23610,1979)
(23810,1997)
(24010,1997)
(24210,2007)
(24410,2007)
(24610,2007)
(24810,2046)
(25010,2107)
  };
\addlegendentry{Edges - Provenance Graph}
\addplot[sharp plot,mark=triangle,GREEN, error bars/.cd, y dir=both, y explicit] plot coordinates
  {
(10,3)
(210,113)
(410,121)
(610,130)
(810,140)
(1010,155)
(1210,160)
(1410,160)
(1610,160)
(1810,162)
(2010,174)
(2210,177)
(2410,177)
(2610,177)
(2810,177)
(3010,182)
(3210,189)
(3410,192)
(3610,192)
(3810,192)
(4010,192)
(4210,192)
(4410,192)
(4610,192)
(4810,192)
(5010,192)
(5210,192)
(5410,192)
(5610,192)
(5810,192)
(6010,192)
(6210,192)
(6410,222)
(6610,229)
(6810,229)
(7010,229)
(7210,229)
(7410,229)
(7610,229)
(7810,229)
(8010,229)
(8210,229)
(8410,229)
(8610,229)
(8810,231)
(9010,238)
(9210,238)
(9410,238)
(9610,238)
(9810,238)
(10010,238)
(10210,238)
(10410,238)
(10610,238)
(10810,238)
(11010,238)
(11210,238)
(11410,238)
(11610,238)
(11810,238)
(12010,238)
(12210,238)
(12410,238)
(12610,238)
(12810,238)
(13010,238)
(13210,242)
(13410,242)
(13610,242)
(13810,242)
(14010,242)
(14210,242)
(14410,242)
(14610,242)
(14810,242)
(15010,242)
(15210,242)
(15410,242)
(15610,242)
(15810,242)
(16010,242)
(16210,242)
(16410,242)
(16610,242)
(16810,249)
(17010,249)
(17210,249)
(17410,249)
(17610,249)
(17810,249)
(18010,249)
(18210,249)
(18410,249)
(18610,249)
(18810,249)
(19010,249)
(19210,249)
(19410,249)
(19610,249)
(19810,249)
(20010,249)
(20210,249)
(20410,249)
(20610,249)
(20810,249)
(21010,249)
(21210,249)
(21410,249)
(21610,249)
(21810,249)
(22010,249)
(22210,249)
(22410,249)
(22610,249)
(22810,249)
(23010,249)
(23210,249)
(23410,249)
(23610,249)
(23810,249)
(24010,249)
(24210,249)
(24410,249)
  };
\addlegendentry{Edges - Process Tree}
\draw[dashed] (4030,20) -- (4030,2107);
\node[anchor=south west] at (4130,20) {dos attack};
    \end{axis}
\end{tikzpicture}
    \caption{Edge Growth}
    \label{fig:prov_edge_grwoth}
  \end{subfigure}
  \begin{subfigure}[t]{.45\textwidth}
    \centering
\begin{tikzpicture}[scale=0.80]
\definecolor{RED}{RGB}{238, 102, 119}
\definecolor{BLUE}{RGB}{68, 119, 170}
\definecolor{GREEN}{RGB}{34, 136, 51}
\definecolor{PURPLE}{RGB}{170, 51, 119}
\definecolor{CYAN}{RGB}{102, 204, 238}
\definecolor{YELLOW}{RGB}{204, 187, 68}
\definecolor{GREY}{RGB}{187, 187, 187}
    \begin{axis}[
        xlabel=Events,
        ylabel=Counts,
        ymin=0,
        ymax=580.8000000000001,
        scaled y ticks = false,
        scaled x ticks = false,
        y tick label style={/pgf/number format/.cd, fixed, fixed zerofill,precision=0},
        x tick label style={/pgf/number format/.cd, fixed, fixed zerofill,precision=0},
        legend style={at={(0.02,0.98)},anchor=north west }]
\addplot[sharp plot,mark=square,RED, error bars/.cd, y dir=both, y explicit] plot coordinates
  {
(10,3)
(210,42)
(410,48)
(610,54)
(810,62)
(1010,73)
(1210,75)
(1410,75)
(1610,75)
(1810,77)
(2010,84)
(2210,86)
(2410,86)
(2610,86)
(2810,86)
(3010,88)
(3210,90)
(3410,92)
(3610,92)
(3810,92)
(4010,92)
(4210,92)
(4410,92)
(4610,92)
(4810,92)
(5010,92)
(5210,92)
(5410,92)
(5610,92)
(5810,92)
(6010,92)
(6210,92)
(6410,113)
(6610,117)
(6810,117)
(7010,117)
(7210,117)
(7410,117)
(7610,117)
(7810,117)
(8010,117)
(8210,117)
(8410,117)
(8610,117)
(8810,119)
(9010,119)
(9210,119)
(9410,119)
(9610,119)
(9810,119)
(10010,119)
(10210,119)
(10410,119)
(10610,119)
(10810,119)
(11010,119)
(11210,119)
(11410,119)
(11610,119)
(11810,119)
(12010,119)
(12210,119)
(12410,119)
(12610,119)
(12810,119)
(13010,119)
(13210,123)
(13410,123)
(13610,123)
(13810,123)
(14010,123)
(14210,123)
(14410,123)
(14610,123)
(14810,123)
(15010,123)
(15210,123)
(15410,123)
(15610,123)
(15810,123)
(16010,123)
(16210,123)
(16410,123)
(16610,123)
(16810,123)
(17010,123)
(17210,123)
(17410,123)
(17610,123)
(17810,123)
(18010,123)
(18210,123)
(18410,123)
(18610,123)
(18810,123)
(19010,123)
(19210,123)
(19410,123)
(19610,123)
(19810,123)
(20010,123)
(20210,123)
(20410,123)
(20610,123)
(20810,123)
(21010,123)
(21210,123)
(21410,123)
(21610,123)
(21810,123)
(22010,123)
(22210,123)
(22410,123)
(22610,123)
(22810,123)
(23010,123)
(23210,123)
(23410,123)
(23610,123)
(23810,123)
(24010,123)
(24210,123)
(24410,123)
  };
\addlegendentry{Vertices - Pseudo Process}
\addplot[sharp plot,mark=star,BLUE, error bars/.cd, y dir=both, y explicit] plot coordinates
  {
(10,0)
(210,52)
(410,87)
(610,100)
(810,103)
(1010,107)
(1210,147)
(1410,162)
(1610,176)
(1810,178)
(2010,180)
(2210,193)
(2410,194)
(2610,195)
(2810,197)
(3010,197)
(3210,197)
(3410,201)
(3610,202)
(3810,209)
(4010,209)
(4210,213)
(4410,213)
(4610,213)
(4810,213)
(5010,213)
(5210,215)
(5410,217)
(5610,217)
(5810,221)
(6010,222)
(6210,225)
(6410,225)
(6610,229)
(6810,303)
(7010,303)
(7210,330)
(7410,330)
(7610,334)
(7810,334)
(8010,336)
(8210,336)
(8410,338)
(8610,338)
(8810,342)
(9010,353)
(9210,353)
(9410,357)
(9610,357)
(9810,360)
(10010,360)
(10210,360)
(10410,360)
(10610,362)
(10810,364)
(11010,368)
(11210,369)
(11410,372)
(11610,372)
(11810,372)
(12010,374)
(12210,377)
(12410,377)
(12610,381)
(12810,382)
(13010,385)
(13210,395)
(13410,395)
(13610,405)
(13810,412)
(14010,413)
(14210,414)
(14410,414)
(14610,417)
(14810,419)
(15010,419)
(15210,421)
(15410,421)
(15610,426)
(15810,426)
(16010,429)
(16210,429)
(16410,429)
(16610,433)
(16810,433)
(17010,438)
(17210,438)
(17410,440)
(17610,440)
(17810,443)
(18010,443)
(18210,443)
(18410,443)
(18610,447)
(18810,447)
(19010,452)
(19210,452)
(19410,455)
(19610,455)
(19810,455)
(20010,459)
(20210,459)
(20410,463)
(20610,463)
(20810,467)
(21010,467)
(21210,469)
(21410,469)
(21610,471)
(21810,471)
(22010,477)
(22210,477)
(22410,487)
(22610,487)
(22810,489)
(23010,489)
(23210,495)
(23410,495)
(23610,495)
(23810,499)
(24010,499)
(24210,501)
(24410,501)
(24610,501)
(24810,507)
(25010,528)
  };
\addlegendentry{Vertices - Provenance Graph}
\addplot[sharp plot,mark=triangle,GREEN, error bars/.cd, y dir=both, y explicit] plot coordinates
  {
(10,4)
(210,68)
(410,74)
(610,80)
(810,88)
(1010,100)
(1210,102)
(1410,102)
(1610,102)
(1810,104)
(2010,112)
(2210,114)
(2410,114)
(2610,114)
(2810,114)
(3010,116)
(3210,118)
(3410,120)
(3610,120)
(3810,120)
(4010,120)
(4210,120)
(4410,120)
(4610,120)
(4810,120)
(5010,120)
(5210,120)
(5410,120)
(5610,120)
(5810,120)
(6010,120)
(6210,120)
(6410,142)
(6610,146)
(6810,146)
(7010,146)
(7210,146)
(7410,146)
(7610,146)
(7810,146)
(8010,146)
(8210,146)
(8410,146)
(8610,146)
(8810,148)
(9010,150)
(9210,150)
(9410,150)
(9610,150)
(9810,150)
(10010,150)
(10210,150)
(10410,150)
(10610,150)
(10810,150)
(11010,150)
(11210,150)
(11410,150)
(11610,150)
(11810,150)
(12010,150)
(12210,150)
(12410,150)
(12610,150)
(12810,150)
(13010,150)
(13210,154)
(13410,154)
(13610,154)
(13810,154)
(14010,154)
(14210,154)
(14410,154)
(14610,154)
(14810,154)
(15010,154)
(15210,154)
(15410,154)
(15610,154)
(15810,154)
(16010,154)
(16210,154)
(16410,154)
(16610,154)
(16810,156)
(17010,156)
(17210,156)
(17410,156)
(17610,156)
(17810,156)
(18010,156)
(18210,156)
(18410,156)
(18610,156)
(18810,156)
(19010,156)
(19210,156)
(19410,156)
(19610,156)
(19810,156)
(20010,156)
(20210,156)
(20410,156)
(20610,156)
(20810,156)
(21010,156)
(21210,156)
(21410,156)
(21610,156)
(21810,156)
(22010,156)
(22210,156)
(22410,156)
(22610,156)
(22810,156)
(23010,156)
(23210,156)
(23410,156)
(23610,156)
(23810,156)
(24010,156)
(24210,156)
(24410,156)
  };
\addlegendentry{Vertices - Process Tree}
\draw[dashed] (4030,20) -- (4030,528);
\node[anchor=south west] at (4130,20) {dos attack};
    \end{axis}
\end{tikzpicture}
    \caption{Vertex Growth}
    \label{fig:prov_vertex_grwoth}
  \end{subfigure}
  
\caption{Provenance vs Resource Interaction Graph Growth (64-bit VM)}
\label{fig:prov_vs_ri}
\end{figure}

\figurename~\ref{fig:prov_vs_ri} compares the growth of the provenance graph versus the resource interaction graph as the number of events grow.  Table \ref{tab:prov-ri-comparison} summarises the size of the respective graphs for this particular example.  The number of vertices for the psuedo process resource interaction graph is 1/3 the provenance graph.  The number of edges is 10\% that of the provenance graph.  In the next section we provide evidence that the pseudo process resource interaction graph grows in a logarithmic fashion which would make the difference more significant for larger log files.

\subsection{Computational Representation Analysis}
\label{sec:comp-analysis}

Based on typical audit logs, we expect the graph's number of nodes to grow slowly compared to the number of events because the first time a node is created as a resource, it is never created again. Nevertheless, there could be scenarios where new resources are constantly being created, resulting in a linear relationship between nodes and events. The growth of the number of edges is expected to be some multiple of the number of nodes. The first time an edge is created between a pair of nodes is very informative. The interactions that update the edge after this are important but less informative.

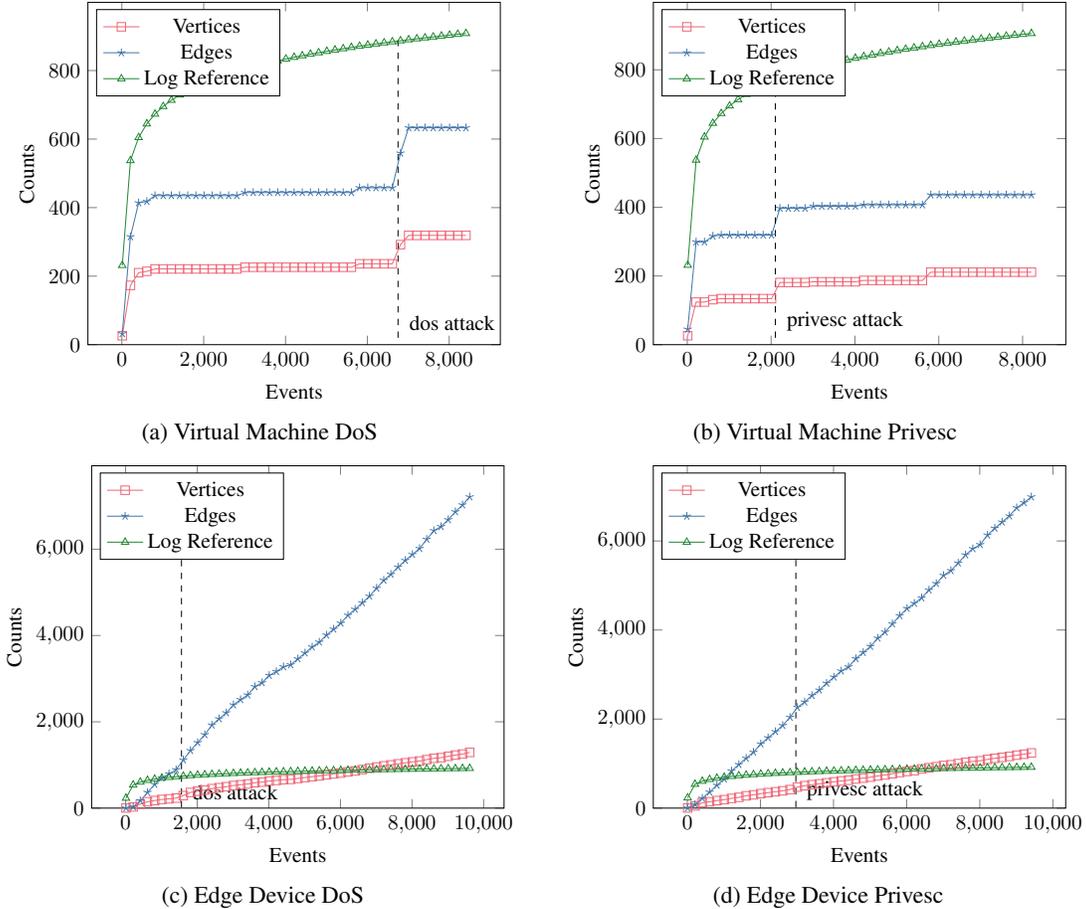
\begin{figure}
  \begin{subfigure}[t]{.45\textwidth}
    \centering
\begin{tikzpicture}[scale=0.80]
\definecolor{RED}{RGB}{238, 102, 119}
\definecolor{BLUE}{RGB}{68, 119, 170}
\definecolor{GREEN}{RGB}{34, 136, 51}
\definecolor{PURPLE}{RGB}{170, 51, 119}
\definecolor{CYAN}{RGB}{102, 204, 238}
\definecolor{YELLOW}{RGB}{204, 187, 68}
\definecolor{GREY}{RGB}{187, 187, 187}
    \begin{axis}[
        xlabel=Events,
        ylabel=Counts,
        ymin=0,
        ymax=999.0516471217243,
        scaled y ticks = false,
        scaled x ticks = false,
        y tick label style={/pgf/number format/.cd, fixed, fixed zerofill,precision=0},
        x tick label style={/pgf/number format/.cd, fixed, fixed zerofill,precision=0},
        legend style={at={(0.02,0.98)},anchor=north west }]
\addplot[sharp plot,mark=square,RED, error bars/.cd, y dir=both, y explicit] plot coordinates
  {
(10,26)
(210,173)
(410,210)
(610,214)
(810,221)
(1010,221)
(1210,221)
(1410,221)
(1610,221)
(1810,221)
(2010,221)
(2210,221)
(2410,221)
(2610,221)
(2810,221)
(3010,226)
(3210,226)
(3410,226)
(3610,226)
(3810,226)
(4010,226)
(4210,226)
(4410,226)
(4610,226)
(4810,226)
(5010,226)
(5210,226)
(5410,226)
(5610,226)
(5810,236)
(6010,236)
(6210,236)
(6410,236)
(6610,236)
(6810,292)
(7010,319)
(7210,319)
(7410,319)
(7610,319)
(7810,319)
(8010,319)
(8210,319)
(8410,319)
  };
\addlegendentry{Vertices}
\addplot[sharp plot,mark=star,BLUE, error bars/.cd, y dir=both, y explicit] plot coordinates
  {
(10,32)
(210,315)
(410,414)
(610,418)
(810,435)
(1010,435)
(1210,435)
(1410,435)
(1610,435)
(1810,435)
(2010,435)
(2210,435)
(2410,435)
(2610,435)
(2810,435)
(3010,444)
(3210,444)
(3410,444)
(3610,444)
(3810,444)
(4010,444)
(4210,444)
(4410,444)
(4610,444)
(4810,444)
(5010,444)
(5210,444)
(5410,444)
(5610,444)
(5810,458)
(6010,458)
(6210,458)
(6410,458)
(6610,458)
(6810,559)
(7010,633)
(7210,633)
(7410,633)
(7610,633)
(7810,633)
(8010,633)
(8210,633)
(8410,633)
  };
\addlegendentry{Edges}
\addplot[sharp plot,mark=triangle,GREEN, error bars/.cd, y dir=both, y explicit] plot coordinates
  {
(10,231.40789255876095)
(210,537.3798730536683)
(410,604.6188059950656)
(610,644.5473072008829)
(810,673.0463887574363)
(1010,695.2236776762827)
(1210,713.380865755909)
(1410,728.7541580653526)
(1610,742.0848177754326)
(1810,753.8525336443056)
(2010,764.3856383560832)
(2210,773.9187679432742)
(2410,782.6254364200635)
(2610,790.6375794340473)
(2810,798.0578615433145)
(3010,804.9677418708857)
(3210,811.432939795501)
(3410,817.5072457703881)
(3610,823.2352443556027)
(3810,828.6543020321274)
(4010,833.7960458529951)
(4210,838.6874818357812)
(4410,843.3518535485758)
(4610,847.8093101100006)
(4810,852.0774322192862)
(5010,856.1716509523379)
(5210,860.105584531646)
(5410,863.8913116244411)
(5610,867.5395950043506)
(5810,871.0600660164339)
(6010,874.4613778102542)
(6210,877.7513334792005)
(6410,880.9369938813468)
(6610,884.0247688894747)
(6810,887.0204950354384)
(7010,889.9295019129195)
(7210,892.7566692368484)
(7410,895.5064760940171)
(7610,898.1830436331878)
(7810,900.7901722162279)
(8010,903.3313738709068)
(8210,905.8099007408166)
(8410,908.2287701106584)
  };
\addlegendentry{Log Reference}
\draw[dashed] (6750,20) -- (6750,908.2287701106584);
\node[anchor=south west] at (6850,20) {dos attack};
    \end{axis}
\end{tikzpicture}
    \caption{Virtual Machine DoS}
    \label{fig:growth_tree_dos_vm}
  \end{subfigure}
  \begin{subfigure}[t]{.45\textwidth}
    \centering
\begin{tikzpicture}[scale=0.80]
\definecolor{RED}{RGB}{238, 102, 119}
\definecolor{BLUE}{RGB}{68, 119, 170}
\definecolor{GREEN}{RGB}{34, 136, 51}
\definecolor{PURPLE}{RGB}{170, 51, 119}
\definecolor{CYAN}{RGB}{102, 204, 238}
\definecolor{YELLOW}{RGB}{204, 187, 68}
\definecolor{GREY}{RGB}{187, 187, 187}
    \begin{axis}[
        xlabel=Events,
        ylabel=Counts,
        ymin=0,
        ymax=996.3908908148984,
        scaled y ticks = false,
        scaled x ticks = false,
        y tick label style={/pgf/number format/.cd, fixed, fixed zerofill,precision=0},
        x tick label style={/pgf/number format/.cd, fixed, fixed zerofill,precision=0},
        legend style={at={(0.02,0.98)},anchor=north west }]
\addplot[sharp plot,mark=square,RED, error bars/.cd, y dir=both, y explicit] plot coordinates
  {
(10,26)
(210,124)
(410,124)
(610,131)
(810,134)
(1010,134)
(1210,134)
(1410,134)
(1610,134)
(1810,134)
(2010,134)
(2210,181)
(2410,181)
(2610,181)
(2810,181)
(3010,183)
(3210,183)
(3410,183)
(3610,183)
(3810,183)
(4010,183)
(4210,187)
(4410,187)
(4610,187)
(4810,187)
(5010,187)
(5210,187)
(5410,187)
(5610,187)
(5810,211)
(6010,211)
(6210,211)
(6410,211)
(6610,211)
(6810,211)
(7010,211)
(7210,211)
(7410,211)
(7610,211)
(7810,211)
(8010,211)
(8210,211)
  };
\addlegendentry{Vertices}
\addplot[sharp plot,mark=star,BLUE, error bars/.cd, y dir=both, y explicit] plot coordinates
  {
(10,45)
(210,299)
(410,299)
(610,316)
(810,319)
(1010,319)
(1210,319)
(1410,319)
(1610,319)
(1810,319)
(2010,319)
(2210,397)
(2410,397)
(2610,397)
(2810,397)
(3010,403)
(3210,403)
(3410,403)
(3610,403)
(3810,403)
(4010,403)
(4210,407)
(4410,407)
(4610,407)
(4810,407)
(5010,407)
(5210,407)
(5410,407)
(5610,407)
(5810,436)
(6010,436)
(6210,436)
(6410,436)
(6610,436)
(6810,436)
(7010,436)
(7210,436)
(7410,436)
(7610,436)
(7810,436)
(8010,436)
(8210,436)
  };
\addlegendentry{Edges}
\addplot[sharp plot,mark=triangle,GREEN, error bars/.cd, y dir=both, y explicit] plot coordinates
  {
(10,231.40789255876095)
(210,537.3798730536683)
(410,604.6188059950656)
(610,644.5473072008829)
(810,673.0463887574363)
(1010,695.2236776762827)
(1210,713.380865755909)
(1410,728.7541580653526)
(1610,742.0848177754326)
(1810,753.8525336443056)
(2010,764.3856383560832)
(2210,773.9187679432742)
(2410,782.6254364200635)
(2610,790.6375794340473)
(2810,798.0578615433145)
(3010,804.9677418708857)
(3210,811.432939795501)
(3410,817.5072457703881)
(3610,823.2352443556027)
(3810,828.6543020321274)
(4010,833.7960458529951)
(4210,838.6874818357812)
(4410,843.3518535485758)
(4610,847.8093101100006)
(4810,852.0774322192862)
(5010,856.1716509523379)
(5210,860.105584531646)
(5410,863.8913116244411)
(5610,867.5395950043506)
(5810,871.0600660164339)
(6010,874.4613778102542)
(6210,877.7513334792005)
(6410,880.9369938813468)
(6610,884.0247688894747)
(6810,887.0204950354384)
(7010,889.9295019129195)
(7210,892.7566692368484)
(7410,895.5064760940171)
(7610,898.1830436331878)
(7810,900.7901722162279)
(8010,903.3313738709068)
(8210,905.8099007408166)
  };
\addlegendentry{Log Reference}
\draw[dashed] (2100,20) -- (2100,905.8099007408166);
\node[anchor=south west] at (2200,20) {privesc attack};
    \end{axis}
\end{tikzpicture}
    \caption{Virtual Machine Privesc}
    \label{fig:growth_tree_privesc_vm}
  \end{subfigure} 
  \vspace{2mm}
  
  \begin{subfigure}[t]{.45\textwidth}
    \centering
\begin{tikzpicture}[scale=0.80]
\definecolor{RED}{RGB}{238, 102, 119}
\definecolor{BLUE}{RGB}{68, 119, 170}
\definecolor{GREEN}{RGB}{34, 136, 51}
\definecolor{PURPLE}{RGB}{170, 51, 119}
\definecolor{CYAN}{RGB}{102, 204, 238}
\definecolor{YELLOW}{RGB}{204, 187, 68}
\definecolor{GREY}{RGB}{187, 187, 187}
    \begin{axis}[
        xlabel=Events,
        ylabel=Counts,
        ymin=0,
        ymax=7933.200000000001,
        scaled y ticks = false,
        scaled x ticks = false,
        y tick label style={/pgf/number format/.cd, fixed, fixed zerofill,precision=0},
        x tick label style={/pgf/number format/.cd, fixed, fixed zerofill,precision=0},
        legend style={at={(0.02,0.98)},anchor=north west }]
\addplot[sharp plot,mark=square,RED, error bars/.cd, y dir=both, y explicit] plot coordinates
  {
(10,4)
(210,27)
(410,94)
(610,147)
(810,176)
(1010,201)
(1210,215)
(1410,232)
(1610,291)
(1810,363)
(2010,391)
(2210,419)
(2410,452)
(2610,475)
(2810,496)
(3010,524)
(3210,542)
(3410,559)
(3610,589)
(3810,603)
(4010,628)
(4210,642)
(4410,659)
(4610,666)
(4810,687)
(5010,708)
(5210,729)
(5410,746)
(5610,771)
(5810,792)
(6010,813)
(6210,841)
(6410,862)
(6610,893)
(6810,924)
(7010,952)
(7210,980)
(7410,1001)
(7610,1026)
(7810,1050)
(8010,1071)
(8210,1092)
(8410,1124)
(8610,1155)
(8810,1169)
(9010,1201)
(9210,1229)
(9410,1253)
(9610,1291)
  };
\addlegendentry{Vertices}
\addplot[sharp plot,mark=star,BLUE, error bars/.cd, y dir=both, y explicit] plot coordinates
  {
(10,3)
(210,32)
(410,171)
(610,364)
(810,551)
(1010,701)
(1210,793)
(1410,890)
(1610,1116)
(1810,1334)
(2010,1518)
(2210,1702)
(2410,1925)
(2610,2070)
(2810,2208)
(3010,2392)
(3210,2514)
(3410,2622)
(3610,2814)
(3810,2906)
(4010,3075)
(4210,3165)
(4410,3274)
(4610,3320)
(4810,3458)
(5010,3596)
(5210,3734)
(5410,3847)
(5610,4010)
(5810,4148)
(6010,4286)
(6210,4470)
(6410,4608)
(6610,4758)
(6810,4912)
(7010,5096)
(7210,5280)
(7410,5418)
(7610,5586)
(7810,5740)
(8010,5878)
(8210,6016)
(8410,6229)
(8610,6430)
(8810,6522)
(9010,6683)
(9210,6867)
(9410,7026)
(9610,7212)
  };
\addlegendentry{Edges}
\addplot[sharp plot,mark=triangle,GREEN, error bars/.cd, y dir=both, y explicit] plot coordinates
  {
(10,231.40789255876095)
(210,537.3798730536683)
(410,604.6188059950656)
(610,644.5473072008829)
(810,673.0463887574363)
(1010,695.2236776762827)
(1210,713.380865755909)
(1410,728.7541580653526)
(1610,742.0848177754326)
(1810,753.8525336443056)
(2010,764.3856383560832)
(2210,773.9187679432742)
(2410,782.6254364200635)
(2610,790.6375794340473)
(2810,798.0578615433145)
(3010,804.9677418708857)
(3210,811.432939795501)
(3410,817.5072457703881)
(3610,823.2352443556027)
(3810,828.6543020321274)
(4010,833.7960458529951)
(4210,838.6874818357812)
(4410,843.3518535485758)
(4610,847.8093101100006)
(4810,852.0774322192862)
(5010,856.1716509523379)
(5210,860.105584531646)
(5410,863.8913116244411)
(5610,867.5395950043506)
(5810,871.0600660164339)
(6010,874.4613778102542)
(6210,877.7513334792005)
(6410,880.9369938813468)
(6610,884.0247688894747)
(6810,887.0204950354384)
(7010,889.9295019129195)
(7210,892.7566692368484)
(7410,895.5064760940171)
(7610,898.1830436331878)
(7810,900.7901722162279)
(8010,903.3313738709068)
(8210,905.8099007408166)
(8410,908.2287701106584)
(8610,910.590786489973)
(8810,912.8985611607154)
(9010,915.1545295303513)
(9210,917.3609665796257)
(9410,919.520000650662)
(9610,921.633625784867)
  };
\addlegendentry{Log Reference}
\draw[dashed] (1560,20) -- (1560,7212);
\node[anchor=south west] at (1660,20) {dos attack};
    \end{axis}
\end{tikzpicture}
    \caption{Edge Device DoS}
    \label{fig:growth_tree_dos_um}
  \end{subfigure}
  \begin{subfigure}[t]{.45\textwidth}
    \centering
\begin{tikzpicture}[scale=0.80]
\definecolor{RED}{RGB}{238, 102, 119}
\definecolor{BLUE}{RGB}{68, 119, 170}
\definecolor{GREEN}{RGB}{34, 136, 51}
\definecolor{PURPLE}{RGB}{170, 51, 119}
\definecolor{CYAN}{RGB}{102, 204, 238}
\definecolor{YELLOW}{RGB}{204, 187, 68}
\definecolor{GREY}{RGB}{187, 187, 187}
    \begin{axis}[
        xlabel=Events,
        ylabel=Counts,
        ymin=0,
        ymax=7689.000000000001,
        scaled y ticks = false,
        scaled x ticks = false,
        y tick label style={/pgf/number format/.cd, fixed, fixed zerofill,precision=0},
        x tick label style={/pgf/number format/.cd, fixed, fixed zerofill,precision=0},
        legend style={at={(0.02,0.98)},anchor=north west }]
\addplot[sharp plot,mark=square,RED, error bars/.cd, y dir=both, y explicit] plot coordinates
  {
(10,4)
(210,49)
(410,121)
(610,143)
(810,169)
(1010,190)
(1210,218)
(1410,242)
(1610,277)
(1810,298)
(2010,326)
(2210,347)
(2410,368)
(2610,389)
(2810,417)
(3010,472)
(3210,504)
(3410,527)
(3610,545)
(3810,569)
(4010,590)
(4210,611)
(4410,625)
(4610,653)
(4810,674)
(5010,695)
(5210,723)
(5410,744)
(5610,772)
(5810,800)
(6010,824)
(6210,842)
(6410,881)
(6610,907)
(6810,930)
(7010,958)
(7210,974)
(7410,1000)
(7610,1028)
(7810,1049)
(8010,1063)
(8210,1095)
(8410,1119)
(8610,1140)
(8810,1161)
(9010,1189)
(9210,1211)
(9410,1240)
  };
\addlegendentry{Vertices}
\addplot[sharp plot,mark=star,BLUE, error bars/.cd, y dir=both, y explicit] plot coordinates
  {
(10,3)
(210,82)
(410,222)
(610,361)
(810,514)
(1010,652)
(1210,836)
(1410,979)
(1610,1120)
(1810,1258)
(2010,1442)
(2210,1580)
(2410,1718)
(2610,1856)
(2810,2040)
(3010,2262)
(3210,2383)
(3410,2529)
(3610,2651)
(3810,2805)
(4010,2943)
(4210,3081)
(4410,3173)
(4610,3357)
(4810,3495)
(5010,3633)
(5210,3817)
(5410,3955)
(5610,4139)
(5810,4323)
(6010,4483)
(6210,4599)
(6410,4722)
(6610,4898)
(6810,5044)
(7010,5228)
(7210,5334)
(7410,5504)
(7610,5688)
(7810,5826)
(8010,5918)
(8210,6132)
(8410,6286)
(8610,6424)
(8810,6562)
(9010,6746)
(9210,6861)
(9410,6990)
  };
\addlegendentry{Edges}
\addplot[sharp plot,mark=triangle,GREEN, error bars/.cd, y dir=both, y explicit] plot coordinates
  {
(10,231.40789255876095)
(210,537.3798730536683)
(410,604.6188059950656)
(610,644.5473072008829)
(810,673.0463887574363)
(1010,695.2236776762827)
(1210,713.380865755909)
(1410,728.7541580653526)
(1610,742.0848177754326)
(1810,753.8525336443056)
(2010,764.3856383560832)
(2210,773.9187679432742)
(2410,782.6254364200635)
(2610,790.6375794340473)
(2810,798.0578615433145)
(3010,804.9677418708857)
(3210,811.432939795501)
(3410,817.5072457703881)
(3610,823.2352443556027)
(3810,828.6543020321274)
(4010,833.7960458529951)
(4210,838.6874818357812)
(4410,843.3518535485758)
(4610,847.8093101100006)
(4810,852.0774322192862)
(5010,856.1716509523379)
(5210,860.105584531646)
(5410,863.8913116244411)
(5610,867.5395950043506)
(5810,871.0600660164339)
(6010,874.4613778102542)
(6210,877.7513334792005)
(6410,880.9369938813468)
(6610,884.0247688894747)
(6810,887.0204950354384)
(7010,889.9295019129195)
(7210,892.7566692368484)
(7410,895.5064760940171)
(7610,898.1830436331878)
(7810,900.7901722162279)
(8010,903.3313738709068)
(8210,905.8099007408166)
(8410,908.2287701106584)
(8610,910.590786489973)
(8810,912.8985611607154)
(9010,915.1545295303513)
(9210,917.3609665796257)
(9410,919.520000650662)
  };
\addlegendentry{Log Reference}
\draw[dashed] (2970,20) -- (2970,6990);
\node[anchor=south west] at (3070,20) {privesc attack};
    \end{axis}
\end{tikzpicture}
    \caption{Edge Device Privesc}
    \label{fig:growth_tree_privesc_um}
  \end{subfigure}

\caption{Process Tree RI Graph Node/Edge Growth vs Audit Events}
\label{fig:growth-tree}
\end{figure}

We hypothesis that attack events will typically generate new edges instead of reinforcing existing ones. The DoS attack involves approximately twice the number of events than the Privesc attack.  Thus we further hypothesis that the DoS attacks will be easier to detect than the Privesc attack.  We take two exemplary audit log files from the virtual machine for the two types of attacks. Note that we omit the first and last two hundred events to avoid the starting and stopping experiment processes. We plot the number of events versus the graph's number of vertices and the number of edges for each file. We also plot the following logarithmic function, determined experimentally to serve as a reference.

\begin{equation}
    \textit{Log Reference} = \log_{1.02}(\textit{\# events})
\end{equation}

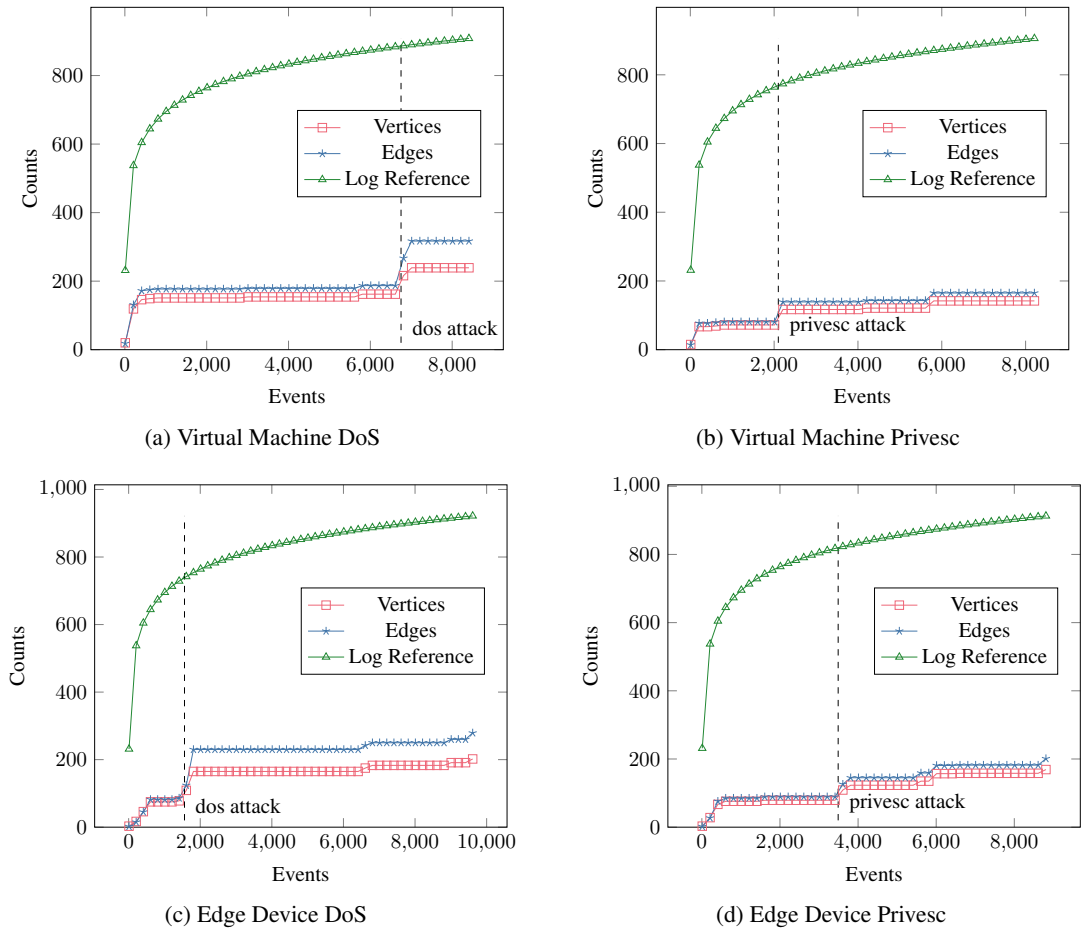
\begin{figure}
  \begin{subfigure}[t]{.45\textwidth}
    \centering
\begin{tikzpicture}[scale=0.80]
\definecolor{RED}{RGB}{238, 102, 119}
\definecolor{BLUE}{RGB}{68, 119, 170}
\definecolor{GREEN}{RGB}{34, 136, 51}
\definecolor{PURPLE}{RGB}{170, 51, 119}
\definecolor{CYAN}{RGB}{102, 204, 238}
\definecolor{YELLOW}{RGB}{204, 187, 68}
\definecolor{GREY}{RGB}{187, 187, 187}
    \begin{axis}[
        xlabel=Events,
        ylabel=Counts,
        ymin=0,
        ymax=999.0516471217243,
        scaled y ticks = false,
        scaled x ticks = false,
        y tick label style={/pgf/number format/.cd, fixed, fixed zerofill,precision=0},
        x tick label style={/pgf/number format/.cd, fixed, fixed zerofill,precision=0},
        legend style={at={(0.50,0.70)},anchor=north west }]
\addplot[sharp plot,mark=square,RED, error bars/.cd, y dir=both, y explicit] plot coordinates
  {
(10,20)
(210,119)
(410,146)
(610,149)
(810,151)
(1010,151)
(1210,151)
(1410,151)
(1610,151)
(1810,151)
(2010,151)
(2210,151)
(2410,151)
(2610,151)
(2810,151)
(3010,154)
(3210,154)
(3410,154)
(3610,154)
(3810,154)
(4010,154)
(4210,154)
(4410,154)
(4610,154)
(4810,154)
(5010,154)
(5210,154)
(5410,154)
(5610,154)
(5810,162)
(6010,162)
(6210,162)
(6410,162)
(6610,162)
(6810,216)
(7010,239)
(7210,239)
(7410,239)
(7610,239)
(7810,239)
(8010,239)
(8210,239)
(8410,239)
  };
\addlegendentry{Vertices}
\addplot[sharp plot,mark=star,BLUE, error bars/.cd, y dir=both, y explicit] plot coordinates
  {
(10,18)
(210,131)
(410,172)
(610,175)
(810,177)
(1010,177)
(1210,177)
(1410,177)
(1610,177)
(1810,177)
(2010,177)
(2210,177)
(2410,177)
(2610,177)
(2810,177)
(3010,179)
(3210,179)
(3410,179)
(3610,179)
(3810,179)
(4010,179)
(4210,179)
(4410,179)
(4610,179)
(4810,179)
(5010,179)
(5210,179)
(5410,179)
(5610,179)
(5810,187)
(6010,187)
(6210,187)
(6410,187)
(6610,187)
(6810,267)
(7010,317)
(7210,317)
(7410,317)
(7610,317)
(7810,317)
(8010,317)
(8210,317)
(8410,317)
  };
\addlegendentry{Edges}
\addplot[sharp plot,mark=triangle,GREEN, error bars/.cd, y dir=both, y explicit] plot coordinates
  {
(10,231.40789255876095)
(210,537.3798730536683)
(410,604.6188059950656)
(610,644.5473072008829)
(810,673.0463887574363)
(1010,695.2236776762827)
(1210,713.380865755909)
(1410,728.7541580653526)
(1610,742.0848177754326)
(1810,753.8525336443056)
(2010,764.3856383560832)
(2210,773.9187679432742)
(2410,782.6254364200635)
(2610,790.6375794340473)
(2810,798.0578615433145)
(3010,804.9677418708857)
(3210,811.432939795501)
(3410,817.5072457703881)
(3610,823.2352443556027)
(3810,828.6543020321274)
(4010,833.7960458529951)
(4210,838.6874818357812)
(4410,843.3518535485758)
(4610,847.8093101100006)
(4810,852.0774322192862)
(5010,856.1716509523379)
(5210,860.105584531646)
(5410,863.8913116244411)
(5610,867.5395950043506)
(5810,871.0600660164339)
(6010,874.4613778102542)
(6210,877.7513334792005)
(6410,880.9369938813468)
(6610,884.0247688894747)
(6810,887.0204950354384)
(7010,889.9295019129195)
(7210,892.7566692368484)
(7410,895.5064760940171)
(7610,898.1830436331878)
(7810,900.7901722162279)
(8010,903.3313738709068)
(8210,905.8099007408166)
(8410,908.2287701106584)
  };
\addlegendentry{Log Reference}
\draw[dashed] (6750,20) -- (6750,908.2287701106584);
\node[anchor=south west] at (6850,20) {dos attack};
    \end{axis}
\end{tikzpicture}
    \caption{Virtual Machine DoS}
    \label{fig:growth_pseudo_dos_vm}
  \end{subfigure}
  \begin{subfigure}[t]{.45\textwidth}
    \centering
\begin{tikzpicture}[scale=0.80]
\definecolor{RED}{RGB}{238, 102, 119}
\definecolor{BLUE}{RGB}{68, 119, 170}
\definecolor{GREEN}{RGB}{34, 136, 51}
\definecolor{PURPLE}{RGB}{170, 51, 119}
\definecolor{CYAN}{RGB}{102, 204, 238}
\definecolor{YELLOW}{RGB}{204, 187, 68}
\definecolor{GREY}{RGB}{187, 187, 187}
    \begin{axis}[
        xlabel=Events,
        ylabel=Counts,
        ymin=0,
        ymax=996.3908908148984,
        scaled y ticks = false,
        scaled x ticks = false,
        y tick label style={/pgf/number format/.cd, fixed, fixed zerofill,precision=0},
        x tick label style={/pgf/number format/.cd, fixed, fixed zerofill,precision=0},
        legend style={at={(0.50,0.70)},anchor=north west }]
\addplot[sharp plot,mark=square,RED, error bars/.cd, y dir=both, y explicit] plot coordinates
  {
(10,15)
(210,67)
(410,67)
(610,69)
(810,72)
(1010,72)
(1210,72)
(1410,72)
(1610,72)
(1810,72)
(2010,72)
(2210,117)
(2410,117)
(2610,117)
(2810,117)
(3010,117)
(3210,117)
(3410,117)
(3610,117)
(3810,117)
(4010,117)
(4210,121)
(4410,121)
(4610,121)
(4810,121)
(5010,121)
(5210,121)
(5410,121)
(5610,121)
(5810,142)
(6010,142)
(6210,142)
(6410,142)
(6610,142)
(6810,142)
(7010,142)
(7210,142)
(7410,142)
(7610,142)
(7810,142)
(8010,142)
(8210,142)
  };
\addlegendentry{Vertices}
\addplot[sharp plot,mark=star,BLUE, error bars/.cd, y dir=both, y explicit] plot coordinates
  {
(10,14)
(210,77)
(410,77)
(610,79)
(810,81)
(1010,81)
(1210,81)
(1410,81)
(1610,81)
(1810,81)
(2010,81)
(2210,139)
(2410,139)
(2610,139)
(2810,139)
(3010,139)
(3210,139)
(3410,139)
(3610,139)
(3810,139)
(4010,139)
(4210,143)
(4410,143)
(4610,143)
(4810,143)
(5010,143)
(5210,143)
(5410,143)
(5610,143)
(5810,165)
(6010,165)
(6210,165)
(6410,165)
(6610,165)
(6810,165)
(7010,165)
(7210,165)
(7410,165)
(7610,165)
(7810,165)
(8010,165)
(8210,165)
  };
\addlegendentry{Edges}
\addplot[sharp plot,mark=triangle,GREEN, error bars/.cd, y dir=both, y explicit] plot coordinates
  {
(10,231.40789255876095)
(210,537.3798730536683)
(410,604.6188059950656)
(610,644.5473072008829)
(810,673.0463887574363)
(1010,695.2236776762827)
(1210,713.380865755909)
(1410,728.7541580653526)
(1610,742.0848177754326)
(1810,753.8525336443056)
(2010,764.3856383560832)
(2210,773.9187679432742)
(2410,782.6254364200635)
(2610,790.6375794340473)
(2810,798.0578615433145)
(3010,804.9677418708857)
(3210,811.432939795501)
(3410,817.5072457703881)
(3610,823.2352443556027)
(3810,828.6543020321274)
(4010,833.7960458529951)
(4210,838.6874818357812)
(4410,843.3518535485758)
(4610,847.8093101100006)
(4810,852.0774322192862)
(5010,856.1716509523379)
(5210,860.105584531646)
(5410,863.8913116244411)
(5610,867.5395950043506)
(5810,871.0600660164339)
(6010,874.4613778102542)
(6210,877.7513334792005)
(6410,880.9369938813468)
(6610,884.0247688894747)
(6810,887.0204950354384)
(7010,889.9295019129195)
(7210,892.7566692368484)
(7410,895.5064760940171)
(7610,898.1830436331878)
(7810,900.7901722162279)
(8010,903.3313738709068)
(8210,905.8099007408166)
  };
\addlegendentry{Log Reference}
\draw[dashed] (2100,20) -- (2100,905.8099007408166);
\node[anchor=south west] at (2200,20) {privesc attack};
    \end{axis}
\end{tikzpicture}
    \caption{Virtual Machine Privesc}
    \label{fig:growth_pseudo_privesc_vm}
  \end{subfigure} 
  \vspace{2mm}
  
  \begin{subfigure}[t]{.45\textwidth}
    \centering
\begin{tikzpicture}[scale=0.80]
\definecolor{RED}{RGB}{238, 102, 119}
\definecolor{BLUE}{RGB}{68, 119, 170}
\definecolor{GREEN}{RGB}{34, 136, 51}
\definecolor{PURPLE}{RGB}{170, 51, 119}
\definecolor{CYAN}{RGB}{102, 204, 238}
\definecolor{YELLOW}{RGB}{204, 187, 68}
\definecolor{GREY}{RGB}{187, 187, 187}
    \begin{axis}[
        xlabel=Events,
        ylabel=Counts,
        ymin=0,
        ymax=1013.7969883633539,
        scaled y ticks = false,
        scaled x ticks = false,
        y tick label style={/pgf/number format/.cd, fixed, fixed zerofill,precision=0},
        x tick label style={/pgf/number format/.cd, fixed, fixed zerofill,precision=0},
        legend style={at={(0.50,0.70)},anchor=north west }]
\addplot[sharp plot,mark=square,RED, error bars/.cd, y dir=both, y explicit] plot coordinates
  {
(10,3)
(210,17)
(410,46)
(610,75)
(810,75)
(1010,75)
(1210,75)
(1410,78)
(1610,109)
(1810,165)
(2010,165)
(2210,165)
(2410,165)
(2610,165)
(2810,165)
(3010,165)
(3210,165)
(3410,165)
(3610,165)
(3810,165)
(4010,165)
(4210,165)
(4410,165)
(4610,165)
(4810,165)
(5010,165)
(5210,165)
(5410,165)
(5610,165)
(5810,165)
(6010,165)
(6210,165)
(6410,165)
(6610,175)
(6810,183)
(7010,183)
(7210,183)
(7410,183)
(7610,183)
(7810,183)
(8010,183)
(8210,183)
(8410,183)
(8610,183)
(8810,183)
(9010,191)
(9210,191)
(9410,191)
(9610,202)
  };
\addlegendentry{Vertices}
\addplot[sharp plot,mark=star,BLUE, error bars/.cd, y dir=both, y explicit] plot coordinates
  {
(10,2)
(210,15)
(410,45)
(610,82)
(810,82)
(1010,82)
(1210,82)
(1410,86)
(1610,124)
(1810,230)
(2010,230)
(2210,230)
(2410,230)
(2610,230)
(2810,230)
(3010,230)
(3210,230)
(3410,230)
(3610,230)
(3810,230)
(4010,230)
(4210,230)
(4410,230)
(4610,230)
(4810,230)
(5010,230)
(5210,230)
(5410,230)
(5610,230)
(5810,230)
(6010,230)
(6210,230)
(6410,230)
(6610,242)
(6810,250)
(7010,250)
(7210,250)
(7410,250)
(7610,250)
(7810,250)
(8010,250)
(8210,250)
(8410,250)
(8610,250)
(8810,250)
(9010,260)
(9210,260)
(9410,260)
(9610,279)
  };
\addlegendentry{Edges}
\addplot[sharp plot,mark=triangle,GREEN, error bars/.cd, y dir=both, y explicit] plot coordinates
  {
(10,231.40789255876095)
(210,537.3798730536683)
(410,604.6188059950656)
(610,644.5473072008829)
(810,673.0463887574363)
(1010,695.2236776762827)
(1210,713.380865755909)
(1410,728.7541580653526)
(1610,742.0848177754326)
(1810,753.8525336443056)
(2010,764.3856383560832)
(2210,773.9187679432742)
(2410,782.6254364200635)
(2610,790.6375794340473)
(2810,798.0578615433145)
(3010,804.9677418708857)
(3210,811.432939795501)
(3410,817.5072457703881)
(3610,823.2352443556027)
(3810,828.6543020321274)
(4010,833.7960458529951)
(4210,838.6874818357812)
(4410,843.3518535485758)
(4610,847.8093101100006)
(4810,852.0774322192862)
(5010,856.1716509523379)
(5210,860.105584531646)
(5410,863.8913116244411)
(5610,867.5395950043506)
(5810,871.0600660164339)
(6010,874.4613778102542)
(6210,877.7513334792005)
(6410,880.9369938813468)
(6610,884.0247688894747)
(6810,887.0204950354384)
(7010,889.9295019129195)
(7210,892.7566692368484)
(7410,895.5064760940171)
(7610,898.1830436331878)
(7810,900.7901722162279)
(8010,903.3313738709068)
(8210,905.8099007408166)
(8410,908.2287701106584)
(8610,910.590786489973)
(8810,912.8985611607154)
(9010,915.1545295303513)
(9210,917.3609665796257)
(9410,919.520000650662)
(9610,921.633625784867)
  };
\addlegendentry{Log Reference}
\draw[dashed] (1560,20) -- (1560,921.633625784867);
\node[anchor=south west] at (1660,20) {dos attack};
    \end{axis}
\end{tikzpicture}
    \caption{Edge Device DoS}
    \label{fig:growth_pseudo_dos_um}
  \end{subfigure}
  \begin{subfigure}[t]{.45\textwidth}
    \centering
\begin{tikzpicture}[scale=0.80]
\definecolor{RED}{RGB}{238, 102, 119}
\definecolor{BLUE}{RGB}{68, 119, 170}
\definecolor{GREEN}{RGB}{34, 136, 51}
\definecolor{PURPLE}{RGB}{170, 51, 119}
\definecolor{CYAN}{RGB}{102, 204, 238}
\definecolor{YELLOW}{RGB}{204, 187, 68}
\definecolor{GREY}{RGB}{187, 187, 187}
    \begin{axis}[
        xlabel=Events,
        ylabel=Counts,
        ymin=0,
        ymax=1004.188417276787,
        scaled y ticks = false,
        scaled x ticks = false,
        y tick label style={/pgf/number format/.cd, fixed, fixed zerofill,precision=0},
        x tick label style={/pgf/number format/.cd, fixed, fixed zerofill,precision=0},
        legend style={at={(0.50,0.70)},anchor=north west }]
\addplot[sharp plot,mark=square,RED, error bars/.cd, y dir=both, y explicit] plot coordinates
  {
(10,3)
(210,28)
(410,67)
(610,76)
(810,76)
(1010,76)
(1210,76)
(1410,76)
(1610,79)
(1810,79)
(2010,79)
(2210,79)
(2410,79)
(2610,79)
(2810,79)
(3010,79)
(3210,79)
(3410,79)
(3610,109)
(3810,123)
(4010,123)
(4210,123)
(4410,123)
(4610,123)
(4810,123)
(5010,123)
(5210,123)
(5410,123)
(5610,135)
(5810,135)
(6010,157)
(6210,157)
(6410,157)
(6610,158)
(6810,158)
(7010,158)
(7210,158)
(7410,158)
(7610,158)
(7810,158)
(8010,158)
(8210,158)
(8410,158)
(8610,158)
(8810,169)
  };
\addlegendentry{Vertices}
\addplot[sharp plot,mark=star,BLUE, error bars/.cd, y dir=both, y explicit] plot coordinates
  {
(10,2)
(210,27)
(410,76)
(610,85)
(810,85)
(1010,85)
(1210,85)
(1410,85)
(1610,89)
(1810,89)
(2010,89)
(2210,89)
(2410,89)
(2610,89)
(2810,89)
(3010,89)
(3210,89)
(3410,89)
(3610,126)
(3810,145)
(4010,145)
(4210,145)
(4410,145)
(4610,145)
(4810,145)
(5010,145)
(5210,145)
(5410,145)
(5610,159)
(5810,159)
(6010,181)
(6210,181)
(6410,181)
(6610,182)
(6810,182)
(7010,182)
(7210,182)
(7410,182)
(7610,182)
(7810,182)
(8010,182)
(8210,182)
(8410,182)
(8610,182)
(8810,201)
  };
\addlegendentry{Edges}
\addplot[sharp plot,mark=triangle,GREEN, error bars/.cd, y dir=both, y explicit] plot coordinates
  {
(10,231.40789255876095)
(210,537.3798730536683)
(410,604.6188059950656)
(610,644.5473072008829)
(810,673.0463887574363)
(1010,695.2236776762827)
(1210,713.380865755909)
(1410,728.7541580653526)
(1610,742.0848177754326)
(1810,753.8525336443056)
(2010,764.3856383560832)
(2210,773.9187679432742)
(2410,782.6254364200635)
(2610,790.6375794340473)
(2810,798.0578615433145)
(3010,804.9677418708857)
(3210,811.432939795501)
(3410,817.5072457703881)
(3610,823.2352443556027)
(3810,828.6543020321274)
(4010,833.7960458529951)
(4210,838.6874818357812)
(4410,843.3518535485758)
(4610,847.8093101100006)
(4810,852.0774322192862)
(5010,856.1716509523379)
(5210,860.105584531646)
(5410,863.8913116244411)
(5610,867.5395950043506)
(5810,871.0600660164339)
(6010,874.4613778102542)
(6210,877.7513334792005)
(6410,880.9369938813468)
(6610,884.0247688894747)
(6810,887.0204950354384)
(7010,889.9295019129195)
(7210,892.7566692368484)
(7410,895.5064760940171)
(7610,898.1830436331878)
(7810,900.7901722162279)
(8010,903.3313738709068)
(8210,905.8099007408166)
(8410,908.2287701106584)
(8610,910.590786489973)
(8810,912.8985611607154)
  };
\addlegendentry{Log Reference}
\draw[dashed] (3490,20) -- (3490,912.8985611607154);
\node[anchor=south west] at (3590,20) {privesc attack};
    \end{axis}
\end{tikzpicture}
    \caption{Edge Device Privesc}
    \label{fig:growth_pseudo_privesc_um}
  \end{subfigure}

\caption{Psuedo Process RI Graph Node/Edge Growth vs Audit Events}
\label{fig:growth-pseudo}
\end{figure}

Figures \ref{fig:growth-tree} and \ref{fig:growth-pseudo} show the plots of the graph's number of nodes and edges versus the number of events for the process graph and pseudo process graph respectively for the two types of attacks. Interestingly, Figures \ref{fig:growth_tree_dos_vm}, \ref{fig:growth_tree_privesc_vm}, and  \ref{fig:growth-pseudo} support the intuition that the attack events result in more vertices and edges.  This can be seen with the dashed line that denotes when the attack starts.  It is also interesting to note that the DoS shows a significant increase as it involves more events than the Privesc attack. The figures clearly show that the graph vertices and edges are growing logarithmically, or less, for these scenarios.  However, Figures \ref{fig:growth_tree_dos_um} and \ref{fig:growth_tree_privesc_um} are clearly growing linearly relative to the number of events.  We found that the edge device had a recurring process, that was not running on the virtual machine, that routinely created short lived processes (only two system calls before stopping).  This results in the linear behaviour.  The pseudo process graph avoids this since its number of nodes is not a function of the number of unique processes.  We note that the use of temporary files can also result in a large number of nodes.  Though we did not see this behaviour in this data set, future work is to address short lived files in a similar fashion to the short lived processes.

The edge attribute growth analysis is more nuanced. If we store a (timestamp, system call type) tuple on edge for every event, the overall edge attribute storage will increase linearly with the number of events. However, using time interval segmentation would result in a time granularity to storage cost trade-off that can be tuned using the $\delta$ time interval.  If the time interval is infinite, the growth is simply the storage cost of a time segment vector $\times$ the number of edges, which we have shown for typical graphs, is bounded by a logarithmic function.


\section{Anomaly Detection Evaluation}
\label{sec:eval}
We first present the link prediction anomaly detection approach and results.  We then present graph clustering approaches to anomaly detection and their results.

\subsection{Evaluation Metric}
\label{sec:metric}

There are numerous metrics to use to evaluate model performance.  A common approach is to use a test set to make an estimate of the model's generalisation accuracy.  We consider our binary classification scenario denoting \textit{abnormal} as the \textit{positive} class and \textit{negative} as the \textit{normal} class.  Given the actual class and the predicted class, the accuracy can be estimated with some combination of the number of \textit{true positive}, \textit{true negative}, \textit{false positive}, and \textit{false negative}, denoted respectfully as $TP$, $TN$, $FP$, and $FN$.  Accuracy can then be calculated as the number correctly predicted divided by the total $(TP+TN) / (TP+TN+FP+FN)$.  However, for imbalanced data, the accuracy can be misleading.  For example, given 99\% of the instances are \textit{normal}, a classifier could trivially always predict \textit{normal} and achieve 99\% accuracy, biased by the high number of $TN$s.  To address this, the \textit{F1} score can be used instead of accuracy since it ignores $TN$s.  Because our data is imbalanced, we choose to primarily use the $F1$ score for evaluation.  It is calculated as the harmonic mean of the precision and recall as follows.

\begin{equation}
    \textit{precision} = \frac{TP}{TP+FP} \qquad
    \textit{recall} = \frac{TP}{TP+FN} \qquad
    \textit{F1} = 2 \times \frac{\textit{precision} \times \textit{recall}}  {\textit{precision} + \textit{recall}}
\end{equation}

\subsection{Anomaly Detection using Graph Autoencoder (GAE)}
\label{sec:linkprediction}

Once we have converted the audit logs into a \graphrep, further processing is dependent on the use case involved. We demonstrate the utility of the \graphrep by using it to detect anomalous events from real-world data collected from an edge device~\cite{CONTAINER_ESCAPE}. We chose to relate the problem to link detection in a graph. For a given \graphrep, the idea is to first train a graph autoencoder on the graph's normal edges and then predict the anomalous edges involved in an attack from the normal edges. The output of the autoencoder is a reconstructed graph with a reconstruction score for each link in the range $[0.0,1.0]$. A score closer to $1.0$ indicates the link is expected between two vertices, given that it was trained on \textit{normal} data.  We consider just one edge vector per edge by setting $\delta$ to the length of the experiments (which is 15 minutes). 

The graph and variational graph autoencoders~\cite{VGAE2016} (GAE) are designed for link prediction. These autoencoders assume only the graph structure and node attributes. Edge attributes are not applicable because this is a link prediction task. We could omit the edge attributes in the \graphrep and lose all-time information.  However, we choose to convert the edge time attributes into node attributes to retain as much information as possible. We denote this converted graph as the \textit{link prediction graph}.

\begin{figure}
    \centering
    \includegraphics[width=0.9\textwidth]{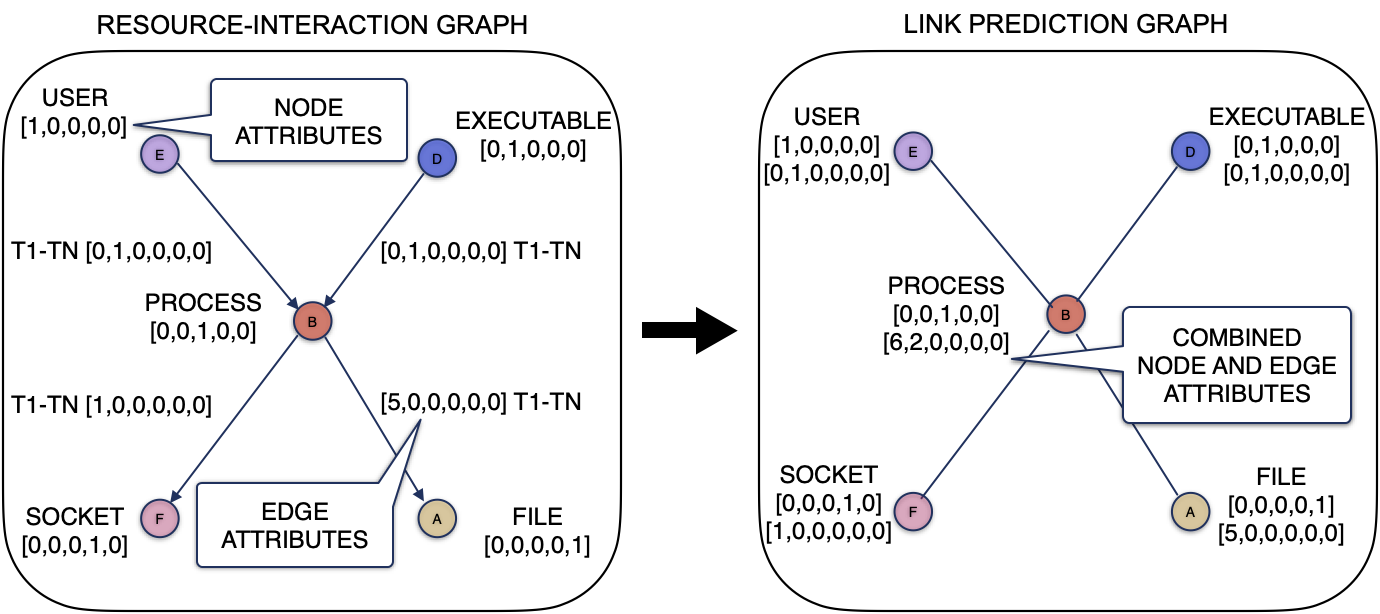}
    \caption{Edge to Node Attributes Conversion}
    \label{fig:usecase-conversion}
\end{figure}

\figurename~\ref{fig:usecase-conversion} depicts an example of converting the edge attributes in a resource-interaction graph into node attributes in the link prediction graph. For simplicity, we denote the edge vector with six entries (i.e. system call type counts). The node type attribute is one-hot encoded as a vector with five entries. For the conversion, each node attribute is appended with edge attributes. Then for each edge, the edge attributes are added to the respective node's edge attributes. As shown, the edge information in the resource-interaction graph is moved into the nodes of the link prediction graph. The edge vector is added twice, once for each node. We also lose the directed edge information as the link prediction graph is undirected.

\begin{table}
    \centering
    \begin{tabular}{l|l|c|c|c}
    \multicolumn{2}{c}{}&\multicolumn{2}{c}{Actual}&\\
    \cline{3-4}
    \multicolumn{2}{c|}{}&Abnormal&Normal&\multicolumn{1}{c}{Total}\\
    \cline{2-4}
    \multirow{2}{*}{Predicted}& Abnormal & $72$ & $20$ & $92$\\
    \cline{2-4}
    & Normal & $8$ & $612$ & $620$\\
    \cline{2-4}
    \multicolumn{1}{c}{} & \multicolumn{1}{c}{Total} & \multicolumn{1}{c}{$80$} & \multicolumn{    1}{c}{$632$} & \multicolumn{1}{c}{$712$}\\
    \end{tabular}
    \caption{Confusion Matrix \\(Threshold=0.30, F1=0.837)}
    \label{tab:privesc-cm}
\end{table}

For each of the auditd logs, we convert them to the \graphrep, then to a link-prediction graph. Before training the graph autoencoder, we perform a random 50\% train/test split for the edges.  The GAE can be configured to use a linear or non-linear encoder.  We used the non-linear encoder because we found that it produced better accuracy.  We train for 10,000 epochs. Table \ref{tab:privesc-cm} shows the confusion matrix for a specific Privesc audit file.  This specific example shows more false positives than false negatives resulting in an F1 score of 0.837 at the optimal threshold of 0.30.


We note that the results are pessimistic because a number of the false negatives may actually be normal edges incorrectly labelled using the time interval annotation approach. Nevertheless, the results demonstrate that, for this example, the approach can effectively identify anomalous edges.

\begin{figure}
  \begin{subfigure}[t]{.45\textwidth}
    \centering
    \begin{tikzpicture}[scale=0.80]
    \begin{axis}[
        xlabel=Number of Epochs,
        ylabel=F1 score,
        grid=major,
        ymin=0.7,
        ymax=1.0,
        scaled y ticks = false, 
        scaled x ticks = false, 
        y tick label style={/pgf/number format/.cd, fixed, fixed zerofill,precision=2},
        x tick label style={/pgf/number format/.cd, fixed, fixed zerofill,precision=0},
        legend pos=south east]

\addplot[sharp plot,mark=square,myred, error bars/.cd, y dir=both, y explicit] plot coordinates {

(200,0.7517907810218326)
(400,0.7739508573921097)
(600,0.7910905574300338)
(800,0.7980968218991297)
(1000,0.8072857504205196)
(1200,0.8160457872747141)
(1400,0.8238863781485913)
(1600,0.828956016933121)
(1800,0.8362526357666841)
(2000,0.8418843174215704)
(2200,0.847967356257025)
(2400,0.8482702514266601)
(2600,0.8511260602031073)
(2800,0.8536644778486624)
(3000,0.8562301431546476)
(3200,0.8555509028272456)
(3400,0.8565638097662778)
(3600,0.8598918483813511)
(3800,0.8582210402654852)
(4000,0.8606440175673677)
(4200,0.8612945417756632)
(4400,0.8612428066372048)
(4600,0.8619647621384061)
(4800,0.8639350928166951)
(5000,0.8639594888635177)
(5200,0.8633804414598641)
(5400,0.8652938453712131)
(5600,0.8673905824921381)
(5800,0.8681385421833744)
(6000,0.8679486292720096)
(6200,0.8681173542685706)
(6400,0.8701508498847832)
(6600,0.8705170890525894)
(6800,0.8704603783978835)
(7000,0.8725524477193771)
(7200,0.8725524477193771)
(7400,0.8726245968083255)
(7600,0.8732788881604286)
(7800,0.8714980207418797)
(8000,0.8716967513383092)
(8200,0.8707019406817665)
(8400,0.8719015628897301)
(8600,0.8712765902008677)
(8800,0.8729384581427745)
(9000,0.8724889783638513)
(9200,0.8719019192609723)
(9400,0.8710666883790209)
(9600,0.8719476199741933)
(9800,0.8740430281275021)
(10000,0.8747314169065005)

    };
\addlegendentry{Privesc-VM}

\addplot[sharp plot,mark=star,myblue, error bars/.cd, y dir=both, y explicit] plot coordinates {

(200,0.8223869517047294)
(400,0.8481836334419762)
(600,0.8625949198172921)
(800,0.8700658399199017)
(1000,0.878990359057011)
(1200,0.8839675752023607)
(1400,0.8880663668204963)
(1600,0.8914896126148706)
(1800,0.8930980069734955)
(2000,0.8961687460116439)
(2200,0.8972349080215583)
(2400,0.8982517211998533)
(2600,0.9000751104901045)
(2800,0.9004983844060676)
(3000,0.9023401771115208)
(3200,0.9041383009134849)
(3400,0.9062668208439706)
(3600,0.9066833459323788)
(3800,0.9081516653591096)
(4000,0.9087917420791068)
(4200,0.9091017972638906)
(4400,0.9094813123153117)
(4600,0.9116310898611667)
(4800,0.9125077778925561)
(5000,0.9136293915861546)
(5200,0.9134136944510526)
(5400,0.9133123612977081)
(5600,0.9139836001288741)
(5800,0.9139836001288741)
(6000,0.9140856181650051)
(6200,0.9147472222795445)
(6400,0.915253738395879)
(6600,0.9153892805645538)
(6800,0.9163017896775641)
(7000,0.9173320284073998)
(7200,0.9191185021821977)
(7400,0.9192341426358117)
(7600,0.9186124457679243)
(7800,0.9188568902494978)
(8000,0.9196584741930588)
(8200,0.9197712935886581)
(8400,0.9190564403640104)
(8600,0.9197874237538698)
(8800,0.9197805557632726)
(9000,0.9202622750702923)
(9200,0.9202264116382416)
(9400,0.9191816205129669)
(9600,0.9204106987448682)
(9800,0.9206180815180405)
(10000,0.9203988245530785)

    };
\addlegendentry{DoS-VM}

\addplot[sharp plot,mark=triangle,mygreen, error bars/.cd, y dir=both, y explicit] plot coordinates {

(200,0.748317062722267)
(400,0.791695756388164)
(600,0.8099847996639864)
(800,0.8273360262347947)
(1000,0.838882061677209)
(1200,0.8477548318204681)
(1400,0.8547299712092647)
(1600,0.8539752110122557)
(1800,0.8592742912724203)
(2000,0.8624883756354997)
(2200,0.8639103563710463)
(2400,0.8653104242602192)
(2600,0.8667081833548969)
(2800,0.868184229052011)
(3000,0.8726596448963686)
(3200,0.8726596448963686)
(3400,0.8721688145485119)
(3600,0.8713961411527462)
(3800,0.873041477077992)
(4000,0.8733059806928747)
(4200,0.874268570735909)
(4400,0.8749344633964031)
(4600,0.8773942977966116)
(4800,0.8792462661179488)
(5000,0.8788426027774192)
(5200,0.8792895785312616)
(5400,0.8798929455748459)
(5600,0.8812387305443521)
(5800,0.8819921606230472)
(6000,0.8843853343666261)
(6200,0.8858051235937924)
(6400,0.8859091901809925)
(6600,0.8856721038210247)
(6800,0.885021401062045)
(7000,0.885021401062045)
(7200,0.885021401062045)
(7400,0.8854270622471642)
(7600,0.8853074831655315)
(7800,0.8826158021223113)
(8000,0.8839601014558239)
(8200,0.8873555793998964)
(8400,0.887841769808118)
(8600,0.8884370079033561)
(8800,0.8884370079033561)
(9000,0.8890733201205507)
(9200,0.8890733201205507)
(9400,0.8890733201205507)
(9600,0.8889990714766072)
(9800,0.8889990714766072)
(10000,0.8889990714766072)

    };
\addlegendentry{Privesc-UM}

\addplot[sharp plot,mark=o,mypurple, error bars/.cd, y dir=both, y explicit] plot coordinates {

(200,0.8260866083925317)
(400,0.846692254295225)
(600,0.8585169992653475)
(800,0.8667873959571575)
(1000,0.8712958909195607)
(1200,0.8724461521677144)
(1400,0.8743555164377725)
(1600,0.8765126915573382)
(1800,0.8763098447486721)
(2000,0.8812170738523945)
(2200,0.8827787886341006)
(2400,0.8837864159503515)
(2600,0.8833321093559733)
(2800,0.8872850543015927)
(3000,0.8882848403587855)
(3200,0.8877442983917075)
(3400,0.8902825734139441)
(3600,0.891876664084332)
(3800,0.893817938084829)
(4000,0.893813357666213)
(4200,0.8939541267831145)
(4400,0.8941485780894617)
(4600,0.894428171980001)
(4800,0.8956383642658103)
(5000,0.8961897089764502)
(5200,0.897268369591603)
(5400,0.8991151532115139)
(5600,0.8991276594643833)
(5800,0.8998152269952466)
(6000,0.900311940494455)
(6200,0.8998105558378594)
(6400,0.9003874759395353)
(6600,0.9021211847465787)
(6800,0.9025118927052749)
(7000,0.9025324861611499)
(7200,0.9039337373158676)
(7400,0.9040733270305785)
(7600,0.9043177073756526)
(7800,0.9049698606917741)
(8000,0.904340822106045)
(8200,0.9047647410354177)
(8400,0.9047647410354177)
(8600,0.9055877484840891)
(8800,0.9060082164395463)
(9000,0.9065669075778401)
(9200,0.9069812699945423)
(9400,0.9070594988196535)
(9600,0.9067103385590162)
(9800,0.9064215886904585)
(10000,0.9062976333911335)

    };
\addlegendentry{DoS-UM}

    \end{axis}
\end{tikzpicture}
    \caption{Pseudo Process Graph}
    \label{fig:accuracy-pseudo}
  \end{subfigure}
  \hfill
  \begin{subfigure}[t]{.45\textwidth}
    \centering
    \begin{tikzpicture}[scale=0.80]
    \begin{axis}[
        xlabel=Number of Epochs,
        ylabel=F1 score,
        grid=major,
        ymin=0.7,
        ymax=1.0,
        scaled y ticks = false, 
        scaled x ticks = false, 
        y tick label style={/pgf/number format/.cd, fixed, fixed zerofill,precision=2},
        x tick label style={/pgf/number format/.cd, fixed, fixed zerofill,precision=0},
        legend pos=south east
        ]

\addplot[sharp plot,mark=square,myred, error bars/.cd, y dir=both, y explicit] plot coordinates {

(200,0.7396906968106625)
(400,0.7604712539254871)
(600,0.7812513579861972)
(800,0.7898939118320875)
(1000,0.800731418371071)
(1200,0.8090049875990595)
(1400,0.8168191494674669)
(1600,0.8180214108289217)
(1800,0.8253446989862874)
(2000,0.8271001697137309)
(2200,0.8341202526798966)
(2400,0.8334736513446145)
(2600,0.8356069599615508)
(2800,0.8368885111554868)
(3000,0.8451958665778311)
(3200,0.8485198557749963)
(3400,0.8503537681921631)
(3600,0.848508145088667)
(3800,0.8506829319528451)
(4000,0.8520260595719971)
(4200,0.8537196659890016)
(4400,0.8535125898398098)
(4600,0.8543683433753375)
(4800,0.8542488690623148)
(5000,0.8564136657018331)
(5200,0.8584084956520022)
(5400,0.8607176087968975)
(5600,0.8620513274056414)
(5800,0.8620928042459526)
(6000,0.8638291317534776)
(6200,0.8638291317534776)
(6400,0.8630885987941509)
(6600,0.8626864392575934)
(6800,0.8631616806570246)
(7000,0.8634374168310931)
(7200,0.8633111595429535)
(7400,0.862926974441086)
(7600,0.863979946638467)
(7800,0.8633911730806593)
(8000,0.8641112344565177)
(8200,0.8669824421639383)
(8400,0.8672964769776135)
(8600,0.868735127740332)
(8800,0.868885156320548)
(9000,0.8704797282849965)
(9200,0.870847380180361)
(9400,0.8722136869035849)
(9600,0.8738355636552583)
(9800,0.8723052060371267)
(10000,0.874166336823511)

    };
\addlegendentry{Privesc-VM}

\addplot[sharp plot,mark=star,myblue, error bars/.cd, y dir=both, y explicit] plot coordinates {

(200,0.8278505353091923)
(400,0.8501094901047288)
(600,0.8613274280269442)
(800,0.8718437026478955)
(1000,0.87708898987812)
(1200,0.882334108794592)
(1400,0.8863986613330823)
(1600,0.8878201135058728)
(1800,0.8910404199625823)
(2000,0.8922794690783661)
(2200,0.8936234310362742)
(2400,0.8958003405517325)
(2600,0.8968952562152737)
(2800,0.8997063729175162)
(3000,0.899860575106131)
(3200,0.9018087808705735)
(3400,0.9015407226420665)
(3600,0.9019494401314413)
(3800,0.9034864974379484)
(4000,0.903947495410997)
(4200,0.90600523979871)
(4400,0.907323991681896)
(4600,0.9070909462275574)
(4800,0.9074709611154208)
(5000,0.9077586883026142)
(5200,0.9092749926753008)
(5400,0.9099256318037128)
(5600,0.9096239487562712)
(5800,0.9095850795848013)
(6000,0.9102336918491682)
(6200,0.9106552726796637)
(6400,0.9102176520355685)
(6600,0.9107742335107303)
(6800,0.9108246362996841)
(7000,0.9111601871867056)
(7200,0.9116395637677455)
(7400,0.9134579335730585)
(7600,0.9141865231750489)
(7800,0.9147928665272657)
(8000,0.9152765009709907)
(8200,0.9155255049550545)
(8400,0.9156550432626588)
(8600,0.916187811595709)
(8800,0.9164034778338842)
(9000,0.9164193099205025)
(9200,0.9161514349044928)
(9400,0.9164557226431876)
(9600,0.9174095945537989)
(9800,0.9173810409670765)
(10000,0.9170991968065325)

    };
\addlegendentry{DoS-VM}

\addplot[sharp plot,mark=triangle,mygreen, error bars/.cd, y dir=both, y explicit] plot coordinates {

(200,0.7731762580677934)
(400,0.7812607159849626)
(600,0.8009246138150179)
(800,0.8135340505792157)
(1000,0.8200114456145219)
(1200,0.8218487311476457)
(1400,0.8314994829462634)
(1600,0.8385386592949707)
(1800,0.8380823375749376)
(2000,0.8392066370756059)
(2200,0.8412273626848426)
(2400,0.8430575861412423)
(2600,0.8452338551504754)
(2800,0.8480193088335962)
(3000,0.8524630512092608)
(3200,0.8525653014025575)
(3400,0.8539489378410676)
(3600,0.8541690764451146)
(3800,0.8543348284563821)
(4000,0.8546729022250421)
(4200,0.85439976635486)
(4400,0.8569029921730436)
(4600,0.8568331691388159)
(4800,0.8563648269648663)
(5000,0.8567044532903043)
(5200,0.8570668888573147)
(5400,0.8571891293510431)
(5600,0.8589757741790084)
(5800,0.8596792373824715)
(6000,0.8587520920889272)
(6200,0.8580998663900508)
(6400,0.8591782753074426)
(6600,0.8590982630918372)
(6800,0.8601960486395308)
(7000,0.8629998131301555)
(7200,0.8627734622682114)
(7400,0.8634468671236015)
(7600,0.8634468671236015)
(7800,0.8641165099807444)
(8000,0.864486946206444)
(8200,0.8670488046517937)
(8400,0.867850789736242)
(8600,0.8672895375860068)
(8800,0.8674001800935566)
(9000,0.8664919038491282)
(9200,0.8664919038491282)
(9400,0.8665956771622438)
(9600,0.8665956771622438)
(9800,0.8666978013452504)
(10000,0.8666978013452504)

    };
\addlegendentry{Privesc-UM}

\addplot[sharp plot,mark=o,mypurple, error bars/.cd, y dir=both, y explicit] plot coordinates {

(200,0.8068226925485955)
(400,0.835314965734065)
(600,0.8488397622876815)
(800,0.8574919554889556)
(1000,0.8612751731986746)
(1200,0.8679659983343035)
(1400,0.8707894893837217)
(1600,0.8749457089698034)
(1800,0.8766413894629134)
(2000,0.8782838257270901)
(2200,0.8806097705758174)
(2400,0.8827648407976043)
(2600,0.8834057628622174)
(2800,0.8856718689074347)
(3000,0.888396722463998)
(3200,0.8901333443554262)
(3400,0.8897886051349627)
(3600,0.8902808966753497)
(3800,0.8909944703694394)
(4000,0.8912144596415835)
(4200,0.8927890467995707)
(4400,0.8946136662334301)
(4600,0.8950755446859204)
(4800,0.8967662580528433)
(5000,0.8977416014905581)
(5200,0.8975066238350594)
(5400,0.8987688236657928)
(5600,0.898659806431712)
(5800,0.898196623336971)
(6000,0.8986456708294511)
(6200,0.8987820717873058)
(6400,0.8997279123844992)
(6600,0.9002252923346831)
(6800,0.9019461187041397)
(7000,0.9018985578255879)
(7200,0.9026072672108785)
(7400,0.9035369990165268)
(7600,0.9034588329316614)
(7800,0.9033731830086473)
(8000,0.9040346060847824)
(8200,0.9047987457341504)
(8400,0.9055315467807431)
(8600,0.9055205546675065)
(8800,0.9070662455133923)
(9000,0.9068020780557563)
(9200,0.9064160402109981)
(9400,0.9049090249327448)
(9600,0.9054754027415516)
(9800,0.9057850554509126)
(10000,0.9055042802074114)

    };
\addlegendentry{DoS-UM}

    \end{axis}
\end{tikzpicture}
    \caption{Process Tree Graph}
    \label{fig:accuracy-tree}
  \end{subfigure}

\caption{Link Prediction Accuracy}
\label{fig:link-prediction-accuracy}
\end{figure}
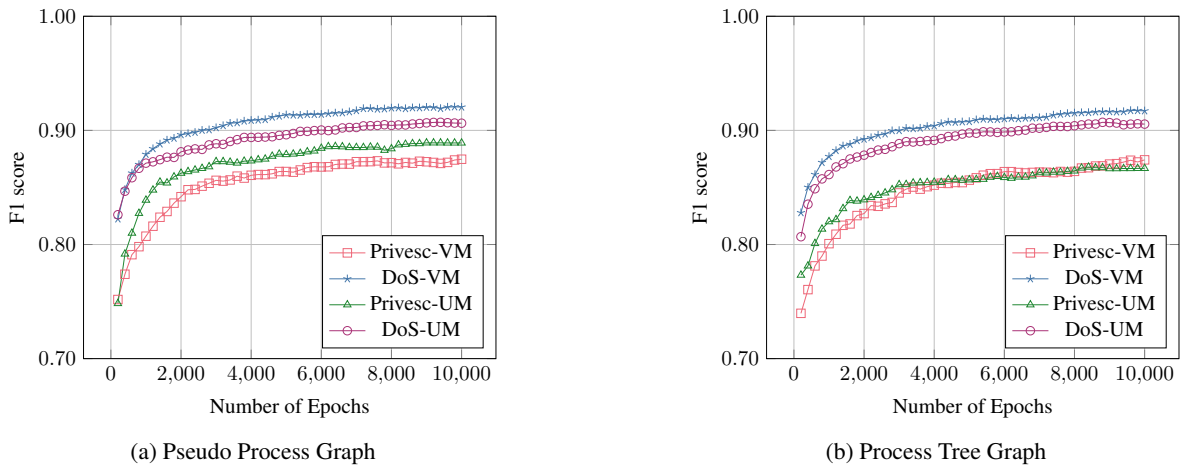

To provide further evidence, we consider 128 audit logs (64 from the edge device and 64 from the virtual machine). Again, we transform the audit log into a graph for each experiment and perform a random 50\% train/test split.  Figures \ref{fig:accuracy-tree} and \ref{fig:accuracy-pseudo} show the respective F1 scores during each epoch.  Both figures support the intuition that the DoS attack is easier to detect than the Privesc attack.  This is more pronounced for the process tree than the pseudo process graph.  The results are roughly equivalent for both the virtual machine and edge device.  Finally, the more compact pseudo process graph appears to get the same accuracy as the larger process tree graph (or even slightly better for the Privesc attack).  This indicates that this compact representation loses little information necessary for this security related task.

\begin{table}
    \centering
    \begin{tabular}[t]{lcc}
    \toprule
    &Node & Node+Edge\\
    \midrule
    Pseudo Process & 0.741 & 0.917 \\
    Process Tree & 0.719 & 0.916 \\
    \bottomrule
    \end{tabular}
    \caption{Node vs Node+Edge Attributes\\(DoS UM, F1 Scores) }
    \label{tab:accruacy-x-vs-xy}
\end{table}

To show the efficacy of the edge to node attribute conversion, we repeat the experiments using only the node attributes and compare to the link prediction graph that uses the node and converted edge attributes.  We use the DoS UM experiments and take the F1 score attained after 10,000 epochs.  Table \ref{tab:accruacy-x-vs-xy} shows the results for both resource interaction graph variants.  In both cases, the node and converted edge attribute representation achieve an accuracy approximately 20\% more than just the node attributes.    



\subsection{Anomaly Detection using Graph Clustering}

To further demonstrate the effectiveness of \graphrep, we select the aforementioned 128 files to experiment with a state-of-the-art graph-based anomaly detector, StreamSpot~\cite{STREAMSPOT}. StreamSpot is a clustering-based detector processing streaming heterogeneous graphs. It vectorizes graphs with features from nodes and edges and uses K-medoids algorithm~\cite{KMedoids} to model benign behaviours. Abnormal graph vectors are detected through comparing their distances from benign clusters. Note that StreamSpot requires each edge in graphs only representing one system call for its graph vectorization approach. Therefore, during constructing \graphrep, we set the hyper-parameter $\delta$ as small as possible to ensure each edge represents only one system call.

For preparation of experiments, we split logs in each scenario (e.g. Dos-UM-Process tree, Dos-VM-Process tree, etc.) into 3 groups (75\% to build training set, 12.5\% to build validating set and 12.5\% to build testing set). To construct training set, we extract all events recorded before attacks in each log. As these events are common tasks running in the containers, they can be regarded as benign data. Note that attacks are uncommon during container running, so they can be detected by models trained with these benign events. The extracted benign events are further converted into resource-interation graphs to build a training set for each scenario. Validating set consists of only benign events and testing set consists of all events including both benign and attack.

\begin{table}
\begin{adjustbox}{width=\columnwidth,center}
\centering
\begin{tabular}{c|ccccccccc}
\hline
\multicolumn{1}{l}{}                                                              & \multicolumn{1}{l}{} & \multicolumn{1}{c}{Accuracy} & \multicolumn{1}{c}{Precision} & \multicolumn{1}{c}{Recall} & \multicolumn{1}{c}{F1-Score} & \multicolumn{1}{c}{Maximum chunk size}  \\ \hline
\multirow{4}{*}{\begin{tabular}[c]{@{}c@{}}Pseudo~\\Process~\\Graph\end{tabular}} 
& Dos-UM               & 0.75~      & 0.67~     & 1.00~     & 0.80~     &44~        \\
& Dos-VM               & 0.75~      & 0.75~     & 0.75~     & 0.75~     &12~        \\
& Privesc-UM           & 0.88~      & 0.80~     & 1.00~     & 0.89~     &28~         \\
& Privesc-VM           & 0.43~      & 0.43~     & 1.00~     & 0.60~     &10~         \\ \hline
\multirow{4}{*}{\begin{tabular}[c]{@{}c@{}}Process~\\Tree~\\Graph\end{tabular}}   
& Dos-UM               & 1.00~      & 1.00~         & 1.00~      & 1.00~          &10~ \\
& Dos-VM               & 0.50~      & 0.50~         & 1.00~      & 0.67~          &10~  \\
& Privesc-UM           & 0.88~      & 0.80~         & 1.00~      & 0.89~          &22~  \\
& Privesc-VM           & 0.71~      & 0.60~         & 1.00~      & 0.75~          &28~   \\ \hline
\end{tabular}
\caption{Evaluation results of StreamSpot. Maximum chunk size is the key hyper-parameter of StreamSpot for constructing graph vectors. We evaluate maximum chunk size with configured values from 10 to 50 and select the best results.}
\label{tab:Evaluation Results of StreamSpot and Unicorn}
\end{adjustbox}
\end{table}

\textbf{Experiment results}
For each dataset, we use 4-fold cross validation to minimise the bias caused by noise within the data and to provide a fair and comprehensive evaluation. The results are shown in Table \ref{tab:Evaluation Results of StreamSpot and Unicorn}.  StreamSpot achieved good detection results, meaning that it can classify majority of attack graphs during the experiments. This is because the \graphrep retains complete attack-related information within edge attributes during construction. With simple graph structure and sufficient attack information, the anomalous subgraph vectors can be identified by StreamSpot.

\section{Related Work}
\label{sec:related_work}

Advanced persistent threats (APT) represent sophisticated attacks that aim to establish an illicit, and long-term presence on a network to extract sensitive data.  Significant research has been conducted using causality graphs to assist in detecting APTs and other intrusions.  King and Chen~\cite{PROV2003} initially proposed employing causality graphs for intrusion analysis.  Lee, et al.~\cite{LOGGC2013}, proposed LogGC, an approach to reduce the size of log files by eliminating events thought to be unnecessary for subsequent forensic analysis.  Hossain, et al.~\cite{Hossain2017}, stored the audit logs into a memory-based, dependency graph data structure by using optimisation techniques to reduce storage requirements. Setayeshfar, et al.~\cite{Setayeshfar2021}, introduced a graphical system for efficiently loading, storing, processing, querying, and displaying system events to support security analysts by compactly storing the logs in multiple formats (main memory, relational database system, hybrid main memory-database system, and a graph-based database). Zhiqiang, et al.~\cite{xu2021depcomm}, provided a graph reduction approach to generate a summary graph from a dependency graph by partitioning a large graph into small process-centric graphs. Zeng, et al.~\cite{Zeng2021}, abstracted behaviours by inferring and aggregating the semantics of audit events (event usage). Yu, et al.~\cite{Yu2021}, combined application (high-level semantics) and audit logs (low-level fine-grained information) to derive a precise attack provenance graph. Winnower~\cite{WINNOWER} proposed a method in which audit records are parsed into graph automata that describes the generic system behaviours of individual nodes in a container cluster.  Sadegh, et al.,~\cite{POIROT2019}, converted auditd logs into a provenance graph to assist in cyber threat detection. Holmes, et al.~\cite{HOLMES2019}, mapped auditd logs to TTP, and detected APT by analysing the correlation between suspicious information flow.  SPADE~\cite{gehani2012spade} provided functionality to collect, store, query and visualise provenance data.  Irshad, et al. \cite{TRACE2021}, leverage SPADE and present TRACE, a provenance tracking system for scalable, real-time, enterprise-wide APT
detection.  Our work utilises the SPADE/TRACE provenance graph representation based on the Open Provenance Model \cite{OPM2011}.

Deep learning for anomaly detection continues to be an active area of research~\cite{DAD2021,DAD2019}.  Autoencoders (AE) are a commonly used technique to learn \textit{normal} data from \textit{abnormal} data for the purposes of anomaly detection.  Separately, research has generalised neural networks, for both supervised and unsupervised scenarios, to taking graphs as input \cite{JURE2016,GCN2016,GIN2018}.  Kipf and Welling \cite{VGAE2016} proposed unsupervised graph autoencoders (GAE) and demonstrated on a link prediction task in citation networks.  Our work leverages the graph autoencoder to predict \textit{normal} \textit{abnormal} edges (i.e. link prediction) on a compressed provenance graph.

More recently, Wang, et al.~\cite{PROVDETECTOR2020}, propose ProvDetector for detecting stealthy malware.  The approach first builds a provenance graph and then reduces the size by only considering certain paths.  The selected paths are then embedded into a vector using \textit{word2vec} for each node in the path to produce a vector, then averaging the vectors.  Finally the embedded path vectors are feed into an unsupervised novelty detection approach based on density (i.e. outliers are treated as an anomaly).



Our work differs, specifically from ProvDetector, in two notable ways.  First and most importantly, we avoid the creation of the provenance graph.  This makes it more suitable for resource-constrained devices (e.g. edge devices).  Second, our anomaly detection approach uses a more recent unsupervised anomaly detection approach based graph neural networks.  To our knowledge this is the first research that combines and analyses compact graph representations with a graph autoencoder (GAE), as well as graph similarity.






\section{Future Work}
\label{sec:future_work}

In this paper we demonstrate that the size of the \graphrep grows slowly with our data set.  Future work will provide a more formal proof given certain assumptions.  This will include other techniques to address short lived files and processes.  Further strengthening the logarithmic growth is also important to mitigate adversarial attacks that could target the anomaly detection system itself.  If the growth is linear then an attacker might be able to cause the graph to grow large enough to exhaust the memory resources of the edge device resulting in an indirect denial of service attack. 

In this work, we compared the process tree and pseudo process graphs. However, future work will also include comparing against, the more elaborate, provenance graphs to exactly determine the extent of performance degradation when using the more compact representation. The data set used is based on container escape scenarios though we submit that it is more generally applicable.  Future work also include using other data sets and/or attack scenarios.

The evaluation approaches required preprocessing of the \graphrep. One potential future area of interest would be to directly leverage the spatial and temporal information without these conversions and evaluate performance along with other metrics (such as computing, and memory footprints etc.) that are relevant for resource constrained deployments.



\section{Conclusion}
\label{sec:conclusion}

Advances in graph clustering and graph neural network techniques are now being applied to various domains.  Cybersecurity has leveraged this to enable novel approaches for analysing operating system audit logs. However, the resulting graphs are often not practical for resource-constrained edge devices. We presented the \graphrep and showed that it is suitable for these devices.  We then showed how the resource-interaction graph could be used for an anomaly detection problem using graph autoencoders and graph clustering techniques. Since both approaches assume no \textit{a priori} attacks, they can potentially detect zero-day attacks.  The results show that the \graphrep can be used to achieve F1 scores typically over 80\% and in some cases over 90\%.

\section{Acknowledgements}
\label{sec:ack}
This work was supported by UK Research and Innovation, Innovate UK [grant number 53707].

\bibliographystyle{unsrt}  
\bibliography{references}

\begin{thebibliography}{10}

\bibitem{POIROT2019}
Sadegh~M. Milajerdi, Birhanu Eshete, Rigel Gjomemo, and V.N. Venkatakrishnan.
\newblock Poirot: Aligning attack behavior with kernel audit records for cyber
  threat hunting.
\newblock In {\em Proceedings of the 2019 ACM SIGSAC Conference on Computer and
  Communications Security}, CCS '19, page 1795–1812, New York, NY, USA, 2019.
  Association for Computing Machinery.

\bibitem{HOLMES2019}
Sadegh~M. Milajerdi, Rigel Gjomemo, Birhanu Eshete, R.~Sekar, and V.N.
  Venkatakrishnan.
\newblock Holmes: Real-time apt detection through correlation of suspicious
  information flows.
\newblock In {\em 2019 IEEE Symposium on Security and Privacy (SP)}, pages
  1137--1152, 2019.

\bibitem{VGAE2016}
Thomas~N. Kipf and Max Welling.
\newblock Variational graph auto-encoders, 2016.

\bibitem{STREAMSPOT}
Emaad Manzoor, Sadegh~M. Milajerdi, and Leman Akoglu.
\newblock Fast memory-efficient anomaly detection in streaming heterogeneous
  graphs.
\newblock In {\em Proceedings of the 22nd ACM SIGKDD International Conference
  on Knowledge Discovery and Data Mining}, KDD '16, page 1035–1044, New York,
  NY, USA, 2016. Association for Computing Machinery.

\bibitem{CONTAINER_ESCAPE}
James Pope and Francesco Raimondo.
\newblock Container escape detection for edge devices.
\newblock \url{https://github.com/jpope8/container-escape-dataset}, 2021.
\newblock Accessed: 2021-09-10.

\bibitem{POPE2021}
James Pope, Francesco Raimondo, Vijay Kumar, Ryan McConville, Rob Piechocki,
  George Oikonomou, Thomas Pasquier, Bo~Luo, Dan Howarth, Ioannis Mavromatis,
  Pietro Carnelli, Adrian Sanchez-Mompo, Theodoros Spyridopoulos, and Aftab
  Khan.
\newblock Container escape detection for edge devices.
\newblock In {\em Proceedings of the 19th ACM Conference on Embedded Networked
  Sensor Systems}, SenSys '21, page 532–536, New York, NY, USA, 2021.
  Association for Computing Machinery.

\bibitem{UMBRELLA_NODE}
BRIL Toshiba~Europe Ltd.
\newblock {UMBRELLA} node.
\newblock \url{https://www.umbrellaiot.com/what-is-umbrella/umbrella-node/},
  2021.
\newblock Accessed: 2021-09-06.

\bibitem{UMBRELLA}
Tim Farnham, Simon Jones, Adnan Aijaz, Yichao Jin, Ioannis Mavromatis, Usman
  Raza, Anthony Portelli, Aleksandar Stanoev, and Mahesh Sooriyabandara.
\newblock Umbrella collaborative robotics testbed and iot platform.
\newblock In {\em 2021 IEEE 18th Annual Consumer Communications Networking
  Conference (CCNC)}, pages 1--7, 2021.

\bibitem{KUBANOMALY}
Chin-Wei Tien, Tse-Yung Huang, Chia-Wei Tien, Ting-Chun Huang, and Sy-Yen Kuo.
\newblock Kubanomaly: Anomaly detection for the docker orchestration platform
  with neural network approaches.
\newblock {\em Engineering Reports}, 1(5):e12080, 2019.

\bibitem{gehani2012spade}
Ashish Gehani and Dawood Tariq.
\newblock Spade: Support for provenance auditing in distributed environments.
\newblock In {\em ACM/IFIP/USENIX International Conference on Distributed
  Systems Platforms and Open Distributed Processing}, pages 101--120. Springer,
  2012.

\bibitem{TRACE2021}
Hassaan Irshad, Gabriela Ciocarlie, Ashish Gehani, Vinod Yegneswaran, Kyu~Hyung
  Lee, Jignesh Patel, Somesh Jha, Yonghwi Kwon, Dongyan Xu, and Xiangyu Zhang.
\newblock Trace: Enterprise-wide provenance tracking for real-time apt
  detection.
\newblock {\em IEEE Transactions on Information Forensics and Security},
  16:4363--4376, 2021.

\bibitem{OPM2011}
Luc Moreau, Ben Clifford, Juliana Freire, Joe Futrelle, Yolanda Gil, Paul
  Groth, Natalia Kwasnikowska, Simon Miles, Paolo Missier, Jim Myers, Beth
  Plale, Yogesh Simmhan, Eric Stephan, and Jan~Van den Bussche.
\newblock The open provenance model core specification (v1.1).
\newblock {\em Future Gener. Comput. Syst.}, 27(6):743–756, jun 2011.

\bibitem{KMedoids}
Leonard Kaufman and Peter~J Rousseeuw.
\newblock {\em Finding groups in data: an introduction to cluster analysis}.
\newblock John Wiley \& Sons, 2009.

\bibitem{PROV2003}
Samuel~T. King and Peter~M. Chen.
\newblock Backtracking intrusions.
\newblock In {\em Proceedings of the Nineteenth ACM Symposium on Operating
  Systems Principles}, SOSP '03, page 223–236, New York, NY, USA, 2003.
  Association for Computing Machinery.

\bibitem{LOGGC2013}
Kyu~Hyung Lee, Xiangyu Zhang, and Dongyan Xu.
\newblock Loggc: Garbage collecting audit log.
\newblock In {\em Proceedings of the 2013 ACM SIGSAC Conference on Computer \&
  Communications Security}, CCS '13, page 1005–1016, New York, NY, USA, 2013.
  Association for Computing Machinery.

\bibitem{Hossain2017}
Md~Nahid Hossain, Sadegh~M. Milajerdi, Junao Wang, Birhanu Eshete, Rigel
  Gjomemo, R.~Sekar, Scott~D. Stoller, and V.~N. Venkatakrishnan.
\newblock {Sleuth: Real-time attack scenario reconstruction from COTS audit
  data}.
\newblock {\em Proceedings of the 26th USENIX Security Symposium}, pages
  487--504, 2017.

\bibitem{Setayeshfar2021}
Omid Setayeshfar, Christian Adkins, Matthew Jones, Kyu~Hyung Lee, and Prashant
  Doshi.
\newblock {GrAALF: Supporting graphical analysis of audit logs for forensics}.
\newblock {\em Software Impacts}, 8, 2021.

\bibitem{xu2021depcomm}
Zhiqiang Xu, Pengcheng Fang, Changlin~Liu Liu, Xusheng Xiao, Yu~Wen, and Dan
  Meng.
\newblock Depcomm: Graph summarization on system audit logs for attack
  investigation.
\newblock In {\em IEEE Symposium on Security and Privacy (SP), San Francisco,
  CA}, pages 22--26, 2021.

\bibitem{Zeng2021}
Jun Zeng, Zheng~Leong Chua, Yinfang Chen, Kaihang Ji, Zhenkai Liang, and Jian
  Mao.
\newblock {WATSON: Abstracting Behaviors from Audit Logs via Aggregation of
  Contextual Semantics}.
\newblock {\em Proceedings 2021 Network and Distributed System Security
  Symposium (NDSS)}, (February), 2021.

\bibitem{Yu2021}
Le~Yu, Shiqing Ma, Zhuo Zhang, Guanhong Tao, Xiangyu Zhang, and Dongyan Xu.
\newblock {ALchemist : Fusing Application and Audit Logs for Precise Attack
  Provenance without Instrumentation}.
\newblock {\em NDSS}, (February), 2021.

\bibitem{WINNOWER}
Wajih~Ul Hassan, Lemay Aguse, Nuraini Aguse, Adam Bates, and Thomas Moyer.
\newblock Towards scalable cluster auditing through grammatical inference over
  provenance graphs.
\newblock In {\em Network and Distributed Systems Security Symposium}, 2018.

\bibitem{DAD2021}
Guansong Pang, Chunhua Shen, Longbing Cao, and Anton Van~Den Hengel.
\newblock Deep learning for anomaly detection: A review.
\newblock {\em ACM Comput. Surv.}, 54(2), mar 2021.

\bibitem{DAD2019}
Raghavendra Chalapathy and Sanjay Chawla.
\newblock Deep learning for anomaly detection: A survey, 2019.

\bibitem{JURE2016}
Aditya Grover and Jure Leskovec.
\newblock Node2vec: Scalable feature learning for networks.
\newblock In {\em Proceedings of the 22nd ACM SIGKDD International Conference
  on Knowledge Discovery and Data Mining}, KDD '16, page 855–864, New York,
  NY, USA, 2016. Association for Computing Machinery.

\bibitem{GCN2016}
Thomas~N. Kipf and Max Welling.
\newblock Semi-supervised classification with graph convolutional networks,
  2016.

\bibitem{GIN2018}
Keyulu Xu, Weihua Hu, Jure Leskovec, and Stefanie Jegelka.
\newblock How powerful are graph neural networks?, 2018.

\bibitem{PROVDETECTOR2020}
Qi~Wang, Wajih Hassan, Ding Li, Kangkook Jee, Xiao Yu, Kexuan Zou, Junghwan
  Rhee, Zhengzhang Chen, Wei Cheng, Carl.A. Gunter, and Haifeng Chen.
\newblock You are what you do: Hunting stealthy malware via data provenance
  analysis.
\newblock In {\em Proc. of the Symposium on Network and Distributed System
  Security (NDSS)}, 01 2020.

\end{thebibliography}

\end{document}